\documentclass[preprint]{aastex}
\author{
M.~Ackermann\altaffilmark{1}, 
M.~Ajello\altaffilmark{2}, 
A.~Allafort\altaffilmark{2}, 
E.~Antolini\altaffilmark{3,4}, 
L.~Baldini\altaffilmark{5}, 
J.~Ballet\altaffilmark{6}, 
G.~Barbiellini\altaffilmark{7,8}, 
D.~Bastieri\altaffilmark{9,10}, 
K.~Bechtol\altaffilmark{2}, 
R.~Bellazzini\altaffilmark{5}
% ...
}
\altaffiltext{1}{Deutsches Elektronen Synchrotron DESY, D-15738 Zeuthen, Germany}
\altaffiltext{2}{W. W. Hansen Experimental Physics Laboratory, Kavli Institute for Particle Astrophysics and Cosmology, Department of Physics and SLAC National Accelerator Laboratory, Stanford University, Stanford, CA 94305, USA}
\altaffiltext{3}{Istituto Nazionale di Fisica Nucleare, Sezione di Perugia, I-06123 Perugia, Italy}
\altaffiltext{4}{Dipartimento di Fisica, Universit\`a degli Studi di Perugia, I-06123 Perugia, Italy}
\altaffiltext{5}{Istituto Nazionale di Fisica Nucleare, Sezione di Pisa, I-56127 Pisa, Italy}
\altaffiltext{6}{Laboratoire AIM, CEA-IRFU/CNRS/Universit\'e Paris Diderot, Service d'Astrophysique, CEA Saclay, 91191 Gif sur Yvette, France}
\altaffiltext{7}{Istituto Nazionale di Fisica Nucleare, Sezione di Trieste, I-34127 Trieste, Italy}
\altaffiltext{8}{Dipartimento di Fisica, Universit\`a di Trieste, I-34127 Trieste, Italy}
\altaffiltext{9}{Istituto Nazionale di Fisica Nucleare, Sezione di Padova, I-35131 Padova, Italy}
\altaffiltext{10}{Dipartimento di Fisica ``G. Galilei", Universit\`a di Padova, I-35131 Padova, Italy}
% ...

\usepackage{graphicx}

\newcommand{\Fermi}{\textit{Fermi}}%
\newcommand{\Xco}{\mbox{$X_\mathrm{CO}$}}

\newcommand{\Wco}{\mbox{$W_\mathrm{CO}$}}
\newcommand{\FGST}{\textit{Fermi Gamma-ray Space Telescope}}
\newcommand{\fermi}{\textit{Fermi}}
\newcommand{\HI}{\mbox{\ion{H}{1}}}%
\newcommand{\NHI}{\mbox{$N(\HI)$}}%
\newcommand{\Hmol}{\mbox{$\mathrm{H}_{2}$}}%
\newcommand{\NHmol}{\mbox{$N(\Hmol)$}}%
\newcommand{\HII}{\ion{H}{2}}%
\newcommand{\Ts}{\mbox{$T_{\mathrm{S}}$}}%
\newcommand{\Av}{\mbox{$A_{\mathrm{V}}$}}%
\newcommand{\EJH}{\mbox{$E(J-H)$}}%

\newcommand{\EBV}{\mbox{$E(B-V)$}}
\newcommand{\EBVres}{\mbox{$E(B-V)_\mathrm{res}$}}

\newcommand{\XcoUnit}{\mbox{$\mathrm{cm^{-2}(K\ km\ s^{-1})^{-1}}$}}

\shorttitle{\fermi\ LAT observation of the Orion region}
\shortauthors{The \fermi\ LAT Collaboration}

\begin{document}


\title{Gamma-ray observations of the Orion Molecular Clouds with the \Fermi\ Large Area Telescope}





\begin{abstract}

We report on the gamma-ray observations of giant molecular clouds
Orion A and B with the Large Area Telescope (LAT)
on-board the \textit{Fermi Gamma-ray Space Telescope}. The gamma-ray emission
in the energy band between $\sim100$~MeV and $\sim100$~GeV is predicted to trace the gas mass distribution in the clouds
through nuclear interactions between the Galactic cosmic rays (CRs) and interstellar gas. The gamma-ray production cross-section for the
nuclear interaction is known to $\sim10$\% precision 
which makes the LAT a powerful tool to measure the gas mass column density
distribution of molecular clouds for a known CR intensity.
We present here such distributions for Orion A and B, and
correlate them with those of the velocity integrated CO intensity (\Wco)
at a $1\arcdeg\times 1\arcdeg$ pixel level.
The correlation is found to be linear over a \Wco\
range of $\sim 10$ fold when divided in 3 regions ,
suggesting penetration of nuclear CRs to most
of the cloud volumes. The \Wco-to-mass conversion factor,
\Xco, is found to be $\sim2.3 \times10^{20}\ \XcoUnit$
for the high-longitude part of Orion A ($l>212\arcdeg$), $\sim 1.7$ times higher than
$\sim 1.3 \times10^{20}$ found for the rest of Orion A and B.
We interpret the apparent high \Xco\ in the high-longitude
region of Orion A in the light of recent works proposing a non-linear
relation between \Hmol\ and CO densities in the diffuse molecular gas.
\Wco\ decreases faster than the \Hmol\ column density in the region
making the gas ``darker'' to \Wco.
\end{abstract}




\section{Introduction}
\label{sec_intro}

The Orion A and B clouds are the archetypes of local giant molecular clouds
(GMCs) where interstellar gas condenses and stars are formed
\citep[e.g.,][and references therein]{Bergin07, Bally08}.
The clouds have been studied
in various wavebands including millimeter observations
of the transition lines between CO rotational states,
especially from $J=1$ to $J=0$
\citep[e.g.,][]{Sanders84, Maddalena86, Dame87, Dame01, Wilson05, Fukui10},
infrared emission \citep[e.g.,][]{IRAS88}, attenuation of
star light \citep[e.g.,][]{Dobashi05}, and near infrared extinction
\citep{Rowles09,Froebrich10,Dobashi11}.
The two clouds are prime targets for the Large Area Telescope (LAT)
on-board the \FGST\ (\Fermi)
in the research of molecular clouds and CR interaction
because they lie isolated from the Galactic plane and no
intense gamma-ray point source overlaps with the clouds
\citep{BrightSrc09,1st_year}.

Gamma rays from the Orion-Monoceros region were first
detected by COS-B in the energy range between 100~MeV and
5~GeV \citep{Caraveo80, Bloemen84}.
EGRET detected gamma rays in the range between 100~MeV and
$\sim 10$~GeV \citep{Digel95, Digel99}. In these studies,
the gamma-ray intensity distribution in a region
including Orion A, B and Monoceros R2 was fitted with
three independent contributions, one proportional to the
atomic hydrogen (\HI) column density, another
proportional to the CO line intensity (\Wco)\footnote{We define \Wco\ as the
velocity-integrated intensity of the transition line between $J=1$ to $J=0$
in $^{12}$C$^{16}$O.}, and the last, a presumed isotropic distribution.
Under the assumptions that \Wco\ traces the \Hmol\ column density,
the CR spectrum doesn't change in the region
and \HI\ spin temperature (\Ts) is constant, the ratio \Xco\ was
determined\footnote{Our \Xco\ is a factor
converting \Wco\ to mass column density measured in units of the proton mass 
in cloud concentrations predominantly consisting of \Hmol.
In some literature \Xco\ is used as the factor converting \Wco\ to \Hmol\ column density.
Where \Wco\ traces \Hmol\ accurately and the chemical state of hydrogen is
predominantly in \Hmol, the 2 definitions are expected to agree. The helium
and heavier atoms are assumed to be mixed uniformly in the interstellar gas
with the solar abundance. We warn readers that comparison of \Xco\ 
values calculated on different CO surveys and gamma-ray observations are not 
straightforward due to differences in their calibration procedure \citep[e.g. see][for the CO calibration factor]{Bronfman88} as well as 
in the assumptions on the CR composition
and the associated cross-sections. }, from the ratio of the gamma-ray intensities
associated with the \HI\ and CO distributions,
to be $\Xco=(2.6\pm1.2)\times 10^{20}\ \XcoUnit$ \citep{Bloemen84}
and $\Xco=(1.35\pm0.15)\times 10^{20}\ \XcoUnit$ \citep{Digel99}.
The ratio was not separately measured for the three clouds,
Orion A, B and Monoceros R2, due to the limited statistics
and spatial resolution of the instruments. We note that
\citet{Strong88} determined \Xco\ on the diffuse
Galactic gamma rays observed by COS-B to be
$\Xco=(2.3\pm0.3)\times 10^{20}\ \XcoUnit$
and \citet{Dame01}, by comparing smoothed infrared intensity
and \Wco\ distributions across the Galaxy, determined it to be
$\Xco=(1.8\pm0.3)\times 10^{20}\ \XcoUnit$.

Since the publications on the EGRET data \citep{Digel95,Digel99},
much progress has been made in studies on Orion A and B:
new observational data became available \citep[e.g.,][]{Dame01, Lombardi01,
Wilson05, Kalberla05, Dobashi05, Rowles09, Froebrich10, Dobashi11}; study of the
molecular clouds was renewed \citep[e.g.,][]{Wilson05,Bally08}; a new modeling of
the Galactic diffuse gamma-ray emission was
proposed incorporating large-scale CR propagation
\citep{Strong98, Strong00};
theoretical calculations of collisional CO rotational-level excitation
were revisited (\citealt{Mengel01,Flower01,Cecchi-Pestellini02, Balakrishnan02,Wernli06,Shepler07};
see also \citealt{Kalberla05, Liszt06, Liszt07})
and the distance to the Orion nebula in the Orion A cloud was measured accurately
\citep{Sandstrom07, Menten07, Hirota07, Kim08}.

The \Fermi\ Gamma-ray Space Telescope mission, launched on 2008 June 11,
has been surveying the sky with the Large Area Telescope (LAT) since 2008 August.
Its wide field of view, large effective area, improved spatial resolution,
and broad energy coverage provide much higher sensitivity
relative to its predecessor EGRET \citep{Atwood09, Calib09}.

Studies based on EGRET observations have established
that gamma rays from Galactic molecular clouds are dominated
by neutral pion decays (which we refer to as the ``pionic gamma rays'' or ``pionic emission'')
in the energy band between 0.2~GeV and 10~GeV
\citep{Bertsch93, Digel95, Digel99}.
Orion A and B are located far ($\sim 8.8$~kpc) from
the Galactic center\footnote{We assume the distance
between the Sun and the Galactic center to be 8.5~kpc and the Galactic rotation velocity
near the Sun to be 220~km~s$^{-1}$.} and displaced from the Galactic plane by $\sim 140$~pc.
The two clouds are only $\sim 400$~pc away from the solar system
where spectra of CR species upto the sub-TeV domain are predicted to be similar to those measured directly at the Earth after correction for the solar modulation.

We can now analyze Orion A and B through the high-energy gamma rays
detected by the \Fermi\ LAT in the light of the recent developments
and study the relation between \Wco\ and mass column density (or \Xco)
in various parts of the Galaxy and obtain the total mass of the clouds\footnote{The mass of Orion A and B is distributed mostly in the column density
range corresponding to a ``translucent'' cloud whose line-of-sight
visual attenuation (\Av) is typically between 1 and 5 mag and has
$n$(\Hmol) typically between 100 and 2000 cm$^{-3}$ \citep[e.g.][]{vanDishoeck88}. }.
The improved spatial resolution and higher gamma-ray statistics
provided by the \Fermi-LAT allow us to determine the relation on angular scales
of $1\times 1$~deg$^2$ (pixels), without being directly affected by the thermodynamical, chemical, or radiation environment inside the Orion clouds, albeit within the limited
angular resolution of the \Fermi\ LAT and uncertainties due to any unresolved weak sources and CR flux variation. The results can be used conversely
to study various environmental effects on \Xco\ in the translucent parts of clouds where
most gas in Orion A and B resides and where the \Xco\ factor has not been straightforward to derive
\citep[e.g.,][]{vanDishoeck86,Magnani88,Bolatto99,Magnani03,Bell06,Snow06,
Bell07,Burgh07,Wall07,Sheffer08}.

Theoretical analyses have long 
suggested that \Xco\ depends on the environment and the \Wco-\NHmol\ relation may be
nonlinear \citep[e.g.,][]{Kutner85, Dickman86, Maloney88, Taylor93, Bolatto99,
Magnani03, Bell07, Burgh07}. Suggestions have also been made that  \Xco\ depends
on the relative abundances of CO, \ion{C}{1}, and \ion{C}{2} \citep[e.g.][]{vanDishoeck88,Hollenbach91,Kopp00}. The existence of gas not traced by \HI\ and CO at the interface between the two phases
 (the ``dark gas'') has been discovered \citep{Grenier05, Planck11}.
The relation between the fraction of carbon in CO
and \Hmol\ density in translucent and diffuse clouds has been updated based on observations and numerical simulations,
for example, by \citet{Burgh10, Wolfire10,Glover10}.
Our results will be interpreted in the light of these recent works.
The \Wco-\NHmol\ relation will be characterized including the ``dark gas,''
and the measured mass column density will be related to the \Av\ value
at which the relation is predicted to become non-linear.

In this paper we analyze diffuse gamma rays spatially associated with the
molecular clouds\footnote{By molecular clouds we mean spatially identified clouds
without distinguishing the small admixture of atomic and ionized hydrogens therein.}
Orion A and B, extract their pionic gamma-ray components,
obtain mass distributions, and compare them with those predicted for \Wco\
measured by \citet{Fukui10} and \citet{Dame01}.
In Section~\ref{sec_data} we describe the
gamma-ray event selection applied in this analysis.
The analysis procedure is described in Section~\ref{sec_ana} in 4 subsections:
the spatial templates used to extract mass column density associated
with multiple emission components are given in Subsection~\ref{subsec_spatial_templates}; energy-binned
spatial fits on the templates are described in Subsection~\ref{subsec_spatial_fit};
the pionic emission is extracted from the spectra obtained in the spatial fits
and \Xco\ is calculated thereon in Subsection~\ref{subsec_spec_ana};
and the total \Hmol\ masses of Orion A and B are estimated in Subsection~\ref{subsec_mass_ana}. In Section~\ref{sec_dis},
we assess systematic uncertainties in the analyses; check the \Xco\
results with recent infrared excess
emission maps by \citet{Dobashi11};
summarize the results; and interpret
them in the light of recent studies of the relation between the \Hmol\ and
CO fraction in the translucent clouds.
The paper is concluded in Section~\ref{sec_conclusion}.

\section{Observations and Data}
\label{sec_data}

The data used in this analysis were obtained in the nominal all-sky survey
mode between 2008 August 4 and 2010 March 11\footnote{Mission Elapsed Time 239,557,413 s through 290,000,000 s where zero is set at 00:00 UTC on 2001 January 1.
During the period, the LAT was operated in the survey mode with the
rocking angle 35~deg (2008 August 4 to 2009 July 9), 39~deg (2009 July 9 to 2009 September 3) and 50~deg (2009 September 3 to 2010 March 11). }.
We select events classified as
\textit{Pass6 Diffuse} class which has a high gamma-ray purity
\citep{Atwood09}. Among the events, we limit the reconstructed zenith angle
to be less than $105\arcdeg$ to greatly reduce gamma rays coming from
the limb of the Earth's atmosphere.
We select the good time intervals (GTIs) of the observations by excluding events that were taken while the instrument rocking angle was larger than $52\arcdeg$.
Another cut is made on the reconstructed gamma-ray energy
at $E_{min} =178$~MeV and $E_\mathrm{max}= 100$~GeV to reduce systematic
uncertainty of the LAT effective area and residual
background events induced by CRs.
Gamma rays in a rectangular region
of $30\arcdeg \times 30\arcdeg$ centered
at ($\ell=210\arcdeg$, $b=-20\arcdeg$) are then selected
for later analyses. We refer to the region as the region-of-interest (ROI) and
the set of events as the data set.

The data set consists of 1,132,436 events of which 901,929 are between 178~MeV and 1~GeV, 224,753 between 1~GeV and 10~GeV, and 5,754
between 10~GeV and 100~GeV. They are binned in $150\times150$ equal-area 
pixels (Hammer-Aitoff projection) in Galactic coordinates with $0.2\arcdeg$ gridding
on their reconstructed arrival directions,
and in 22 logarithmic bins between
$E_\mathrm{min} = 178$~MeV and $E_\mathrm{max}=100$~GeV on their
reconstructed energies.

\begin{figure}
\epsscale{.65}
\plotone{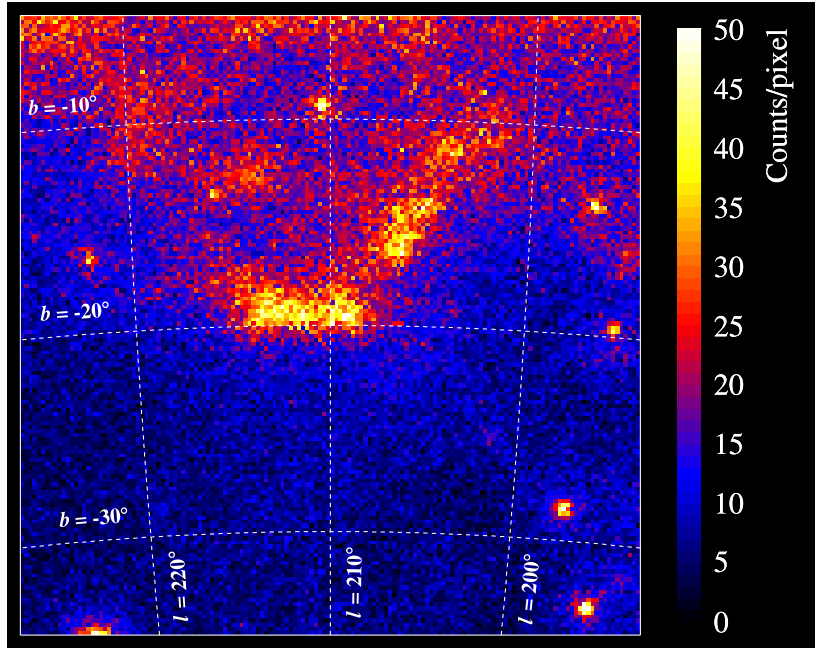}
\caption{Gamma-ray count distribution in the Orion region in the energy band
between 178~MeV and 100~GeV in the Hammer-Aitoff projection
on the Galactic coordinates.
The pixel size is $0.2\times 0.2$~deg$^2$.}
\label{fig_cmap}
\end{figure}

The map of counts integrated over the energy range of the data set is shown in Fig.~\ref{fig_cmap}.
We can visually identify Orion A and B
near the center of the region and the outer Galactic plane in the upper part.
We note that Monoceros R2 is also visible between Orion A/B
and the outer Galactic plane.

\section{Analyses}
\label{sec_ana}

The analyses presented here begin by finding the relationship between the spatial distributions of gamma rays and \Wco, the
most widely used proxy of \Hmol, in the Orion clouds and by studying 
the proportionality between the two and its spatial dependence 
within the Orion clouds.  The analyses proceed in 3 steps.

In the first step, the spatial distribution of the ``background'' gamma rays, i.e., the gamma rays not associated with the \Hmol\  clouds, is determined by using spatial
distribution templates, for the \HI\  gas,
for the inverse Compton scattering (IC) component, for the point sources,
and for a presumed isotropic component (Subsections~\ref{subsec_spatial_templates}).
We then fit, in Subsection~\ref{subsec_spatial_fit}, the gamma-ray spatial
distribution in each of the 22 energy bins as a sum of the ``background'' distribution
and a distribution tentatively associated with the \Hmol\  gas (\Hmol-template).
The ``background'' is subtracted from the measured gamma-ray intensity distribution
and the remainder is defined as the gamma-ray intensity distribution associated with the \Hmol\ gas with which \Wco\ is correlated pixel-by-pixel. We note that    
the gamma-ray intensity measures the mass column density in the \Hmol\ gas 
for a known CR spectrum. We repeat the fit with 2 alternative \Hmol-templates.

In the second step (Subsections~\ref{subsec_spec_ana} and \ref{subsec_mass_ana}),
the energy-binned gamma-ray emissivity for the \Hmol\ gas ($B_i$ in eq.~(\ref{eq_emission})) are
assembled as the gamma-ray spectrum for each of the 3 \Hmol-templates.
The spectrum is then fitted as a sum of the gamma rays produced in the pionic and bremsstrahlung processes.

In the third step, the gamma-ray intensity distribution associated with the pionic emission
is converted to the mass column density.  The \Wco-mass conversion 
factor (\Xco) is calculated via two methods, one
by comparing the gamma-ray counts associated with the \HI\ gas and with the \Hmol\ gas
(the \Hmol/\HI\ method) and the other by dividing the gamma-ray counts of the pionic emission by the number of pionic gamma rays expected per unit gas mass
(the pionic method). In the first method, we assume the CR spectrum is uniform 
in the local \HI\ region within Galactocentric radius of $8-10$~kpc 
(see Subsection~\ref{HI_template}) and in the Orion clouds. In the latter method, 
we assume the CR spectrum including its absolute flux 
is known in the Orion clouds. We validate these assumption using GALPROP. 

We use GALPROP \citep{Strong98, Strong00} with the parameter set
labeled as GALDEF 54\_77Xvarh7S. This parameter set is known to
reproduce reasonably well the Galactic diffuse gamma-ray emission observed
with the LAT \citep{NonGeVExcess}\footnote{A detailed description
of GALDEF files can be found at http://galprop.stanford.edu.}.
We refer to the results obtained by running GALPROP with this parameter set as
the GALPROP results in this paper.

\subsection{Spatial distribution templates}
\label{subsec_spatial_templates}

Initially we assume the gamma-ray emission from the ROI to be made of 4
``background'' components and one ``signal'' \Hmol\ component, each emitting
gamma rays with a characteristic spatial distribution.
The 4 ``background'' components are spatially
associated with the diffuse \HI\ gas, the inverse Compton (IC) scattering
by electrons\footnote{We refer to electrons as a sum of $e^+$ and $e^-$.}
off interstellar radiation fields, the point sources, and the sum of extragalactic
diffuse emission (including unresolved sources) and 
backgrounds induced by CRs in the instrument. 
We assume the last sum to be isotropic
and refer to it as the isotropic component.  We ignore the contribution
from ionized hydrogen gas (\HII) because its density
is low when averaged in $1 \times 1$~deg$^2$ pixels
($< 0.5 \mathrm{cm}^{-3}$) and its total mass
is negligible in the ROI \citep{Gordon69, O'Dell01}. 

All spatial components except for the IC component are assumed to have, individually,
an energy-independent underlying spatial distribution in Galactic
coordinates $(l,b)$.  Another important underlying assumption
is that the nuclear CR spectrum is uniform over the ROI.
We make spatial templates for the 22 energy
bins by convolving the spatial distributions with the energy-dependent
point spread function (PSF) and exposure for the individual energy bins.
Hence the spatial templates are energy dependent. In 
Subsection~\ref{subsec_spec_ana}
we will show that the spectra of the gamma-ray emissions associated with
the \HI\ and \Hmol\ gas consist of the pionic and bremsstrahlung
components.

The gamma-ray intensity $I_\gamma(l, b)$ for the $i$-th energy bin is 
interpreted as the sum of the five contributions, 
each being the product of the normalization factor
for the $i$-th energy bin and the spatial template.  

\begin{equation}
  I_{\gamma,i}(l, b) = A_i N(\HI)(l,b) + B_i N(\Hmol)(l,b) +
     IC_i(l,b) + \sum_j(C_{ij}\delta_{l_j,b_j}) + D_i,
\label{eq_emission}
\end{equation}

The normalization factors are: $A_i$ for the \HI\ gas; $B_i$
for the component associated with clouds consisting predominantly of \Hmol;
$IC_i$ for the inverse Compton component; $C_{ij}\delta_{l_j,b_j}$
for the $j$-th point source at $(l_j, b_j)$; and $D_i$
for the isotropic component which is assumed not to depend on $(l,b)$.
The normalization factors are determined independently for the 22 energy
bins. We note that $IC_i$ are fixed at the values given by GALPROP, because
the spatial distribution is highly correlated with the isotropic component,
and the IC component is sub-dominant in the ROI.

Later in Subsection~\ref{subsec_spatial_fit}, we will explore 3 templates for
\Hmol, two based on \Wco\ and one on \Wco\ plus the ``dark gas'' proposed by \citet{Grenier05}.

\subsubsection{Diffuse \HI\ gas template}\label{HI_template}

Atomic hydrogen gas (\HI) is broadly distributed in the Galaxy with a total
mass exceeding that of molecular hydrogen (\Hmol) \citep[e.g.,][]{Ferriere01, Snow06}.
In the outer Galaxy where the Orion clouds are located, the
mass column density of \HI\ is lower than that of \Hmol\ at the Orion clouds
\citep{Kalberla10, Kalberla05}

We used the Leiden/Argentine/Bonn (LAB) survey data \citep{Kalberla05}
corrected for optical thickness by adopting a constant spin temperature (\Ts)
of 125~K as the \HI\ gas spatial distribution template (see Fig.~\ref{fig_template}a).
The LAB intensity distribution is divided into five annuli
centered at the Galactic center as has been done in other
\Fermi\ diffuse emission analyses \citep{Cassiopeia}.
Their inner and outer Galactocentric radii ($R$) are: 8 to 10, 10 to 11.5,
11.5 to 16.5, 16.5 to 19, and 19 to 50 kpc.
The line-of-sight velocity distribution of the \HI\ gas in the Orion region overlaps
that of the CO gas associated with the Orion clouds and that of the local \HI\ annulus
($R = 8 -10$~kpc) quite well.

Gamma-ray contributions from all the \HI\ annuli overlapping
our ROI have been included in the analyses. In the fitting, the CR intensity
is treated independently at each annulus.
The contributions from annuli other than the local one ($R = 8-10$~kpc)
are through the periphery of the LAT PSF and less than $\sim 5$\% in gamma-ray counts.
Hence our analyses are insensitive to variation in the CR intensity and/or spectrum
among the neighboring annuli.

The spin temperature of \HI\ gas, \Ts, is not well constrained in the region
nor known to be uniform over the ROI: its quoted value in the literature ranges
between $\sim 90$~K and $\sim 400$~K \citep[e.g.][]{Mohan04a, Mohan04b}.
We estimate, later in this paper, the contribution to the overall systematic error
from this uncertainty by repeating the analysis for $\Ts =250$~K  and $90$~K.
No significant concentration of cold \HI\
is known around Orion A and B at large-scale ($>1 \times 1$~deg$^2$)
\citep{Kalberla10}. An exploratory study of cold \HI\ mixed in selected
\Hmol\ cloud cores has found the mean \HI\ fraction to be less than
0.5\% \citep{Krco08}. So we can safely ignore such a mixture in the analysis.

Gamma rays are produced in the \HI\ gas through the pionic and
bremsstrahlung processes with intensities proportional to the
CR nuclear and electron spectra in the gas, respectively.

\subsubsection{Molecular cloud template}
\label{subsubsection_H2templates}
We try 3 \Hmol\ templates to represent the \Hmol\ spatial distribution in the ROI.
In making the templates we assume that
the \Hmol\ column density is proportional to \Wco\
measured by two CO ($J=1 \rightarrow 0$) surveys, one from
NANTEN \citep{Fukui10} covering
the  areas around the Orion clouds
with effective resolution of $4\arcmin$ and the other being
the Galactic survey by \citet{Dame01} covering the ROI with angular
resolution of $8.7\arcmin$. The spatial distributions indicated by the two surveys are
mutually consistent at the angular scale of the LAT PSF except for the overall
normalization.

\begin{figure}
\epsscale{1.0}
\plotone{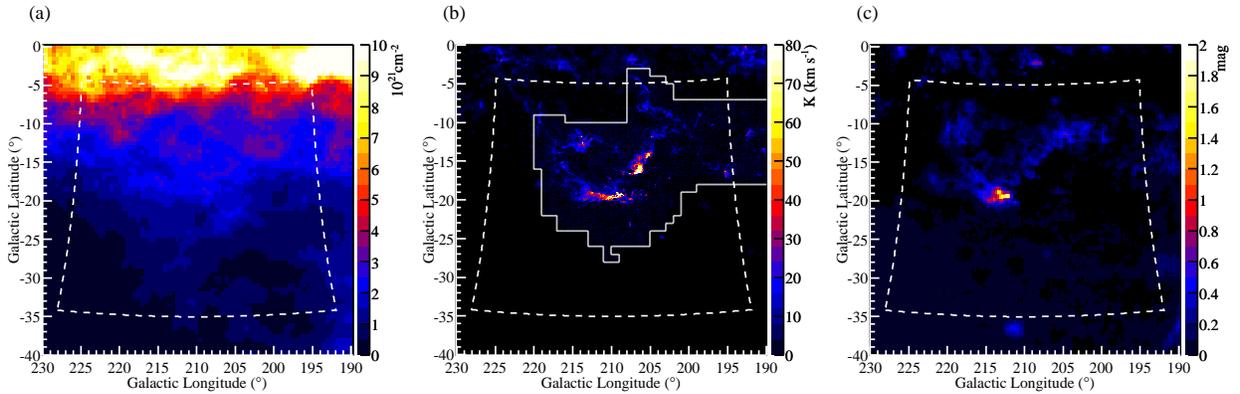}
\caption{(a) \NHI\ template summed over the line-of-sight velocity. The pixel size is $0.5^\circ\times0.5^\circ$. The dashed lines show the boundary of the ROI.
(b) \Wco\ template used in \Hmol-template-1 and \Hmol-template-2.
We used NANTEN data \citep{Fukui10} in the area bounded by the solid lines
and those by \citet{Dame01} elsewhere.
Pixel resolution is $0.125^\circ\times0.125^\circ$. (c) \EBVres\ template used
in \Hmol-template-3. Pixel resolution is $0.5^\circ\times0.5^\circ$.}
\label{fig_template}
\end{figure}

The first \Hmol\ template, \Hmol-template-1, is made
by combining the two surveys and accounting for their relative intensity scales (Fig.~\ref{fig_template}b):
NANTEN \Wco\ for the area defined by the solid white line and 
that by \citet{Dame01} for the rest of the region.
We refer to the 3 regions defined by dashed lines in Fig.~\ref{fig_model_map}b
as ``the 3 Orion regions'' hereafter\footnote{The boundaries are: Orion A Region I ($217\arcdeg > \ell > 212\arcdeg$, $-23\arcdeg < b < -16\arcdeg$),
Orion A Region II ($212\arcdeg > \ell > 205\arcdeg$, $-23\arcdeg < b < -16\arcdeg$, excluding the overlap with Orion B), and Orion B ($209\arcdeg > \ell > 203\arcdeg$, $-18\arcdeg < b < -13\arcdeg$)}\label{footnote_regions}.
We scale the NANTEN data by the factor $1/1.11$ to adjust the intensities to a common scale with \citet{Dame01} because the survey by \citet{Dame01}
has been widely used in gamma-ray analyses.

We first start the analyses by setting one common $B_i$ factor for \Wco\
in the ROI (\Hmol-template-1), or equivalently, one common \Xco\ for the entire ROI.
In the second \Hmol\ template, \Hmol-template-2, the \Wco\ distribution is divided
into 4 regions (the 3 Orion regions and the rest of the ROI)
and allow $B_i$, or equivalently \Xco, to be different in
each region. We add a ``dark gas'' template \citep{Grenier05}
to \Hmol-template-1 to make the third \Hmol\
template, \Hmol-template-3 (Fig.~\ref{fig_template}a and \ref{fig_template}c).
The normalization is set free for the 2 templates.

These spatial templates are described further in the subsections to follow.

\subsubsection{Inverse Compton template}

The inverse Compton component is known to be minor around the Orion clouds.
We use the IC spectrum and spatial distribution given by GALPROP
where the interstellar photon fields are taken from \citet{Porter08}.
The typical Galactic-scale IC intensity in the region is $\sim 5$ times
smaller than the isotropic component described later, and their spatial and spectral distributions are similar in this region.
Possible local enhancement is the IC emission around the Orion Nebula (M42)
where strong ultraviolet emission \citep[e.g.][]{Murthy05} and
moderate infrared emission \citep[e.g.][]{Prisinzano08} exists.
According to our calculation, such IC emissions are not detectable   
with the current LAT sensitivity \citep{Orlando08}.

\subsubsection{Point sources in the Orion region}\label{subsec_pointsrc}

More than 1400 point sources are reported in the First \Fermi\ LAT Catalog \citep{1st_year}. Among them, 30 point sources are
in our ROI, $(l, b) = (210\pm15\arcdeg, -20\pm15\arcdeg)$.
There are an additional 29 sources within 5~deg of the ROI.
In the likelihood fit to be discussed later, the normalization
is set free, energy-bin by energy-bin, for 25 high-confidence sources in the ROI;
the indexes and normalizations are fixed to the values given
in the First \Fermi\ LAT Catalog \citep{1st_year} for those outside of the region.
There are 5 low-confidence sources (or candidates) overlapping
with the clouds: they are\footnote{No new sources have been added in this region in 
the Second \Fermi\ LAT Catalog \citep{2FGL}.}
1FGL J0540.4$-$0737c, J0536.2$-$0607c, J0534.7$-$0531c, J0541.9$-$0204c,
and J0547.0+0020c. 
Their fluxes are all low and labeled as ``c'' in the catalog, meaning
either their flux estimates are uncertain,
or they can be artifacts resulting from incorrect modeling of
the Galactic diffuse emission.
We fit the spatial templates and analyze
the spectra in the 3 Orion regions with and without them. The results we quote
will be for the analyses without them: we include their possible contribution
in the systematic error.

\subsubsection{Isotropic component}

In the present analyses, the extragalactic emission and
residual CR background in the data are not separated
but treated as a single isotropic component \citep{LocalHI09, Cassiopeia,
ThirdQuadrant, EGB}.
The total flux of the component at $1$~GeV is $\sim 25$\% of that
associated with \Hmol\ when averaged over the 3 Orion regions (subtending $\sim30$~msr)
defined around Orion A and B (see Fig.~\ref{fig_model_map}b).

The residual background in the \textit{Pass6 Diffuse} class
consists of CR-induced events misclassified as gamma rays and
CRs that converted in the passive material just outside of the LAT
without leaving a signal in the anti-coincidence detector \citep{Atwood09}.
When averaged over many orbits of observations, the
residual background can be approximated as isotropic.

\subsection{Fit to the Spatial Distribution}
\label{subsec_spatial_fit}
All spatial templates described in the previous subsection were
convolved with the LAT exposure and PSF.
The spatial fit is made using the binned likelihood program \textit{gtlike}
included in the \Fermi\ ScienceTools\footnote{We use ScienceTools version v9r16p0 with \textit{P6\_V3\_DIFFUSE} instrument response functions.}
and the 4 normalizations ($A_i$, $B_i$, $C_{ij}$ and $D_j$)
in Eq.~(\ref{eq_emission}) are determined independently for the 22 energy bins.
We note again that $IC_i$ are fixed at the values given by GALPROP. 
Each \HI\ annulus has a separate $A_i$. We report only $A_i$ for the local 
annulus as others are not determined well because 
they lie mostly outside of our ROI.

Our scientific interest is to study the contributions from the gas concentrations
identified as Orion A and B, which are believed to be predominantly \Hmol.
We consider, hence, the sum of the \HI, IC, point-source,
and isotropic components as the ``background'' which is determined by fitting
the observed gamma-ray distribution for each of the 22 energy bins.
In the fits, we assume that \Hmol-template-1, or the \Wco\ distribution, represents
approximately the \Hmol\ distribution.
The gamma-ray distribution associated with the \Hmol\ gas can be extracted  
less dependently on yet-unknown \Hmol-\Wco\ relation by subtracting the ``background'' 
from the observed gamma-ray distribution.

We define 2 improved \Hmol-templates, \Hmol-template-2 and 3
after the initial analysis on \Hmol-template-1. The spatial distribution is not
proportional to \Wco\ for the 2 improved templates and hence the ``background''
is different for each \Hmol-template by a small amount.
The difference is however negligible.

\subsubsection{Spatial Fit with \Wco\ of One \Xco: \Hmol-template-1}
\label{subsubsec_HiCo}

\begin{figure}
\epsscale{0.9}
\plotone{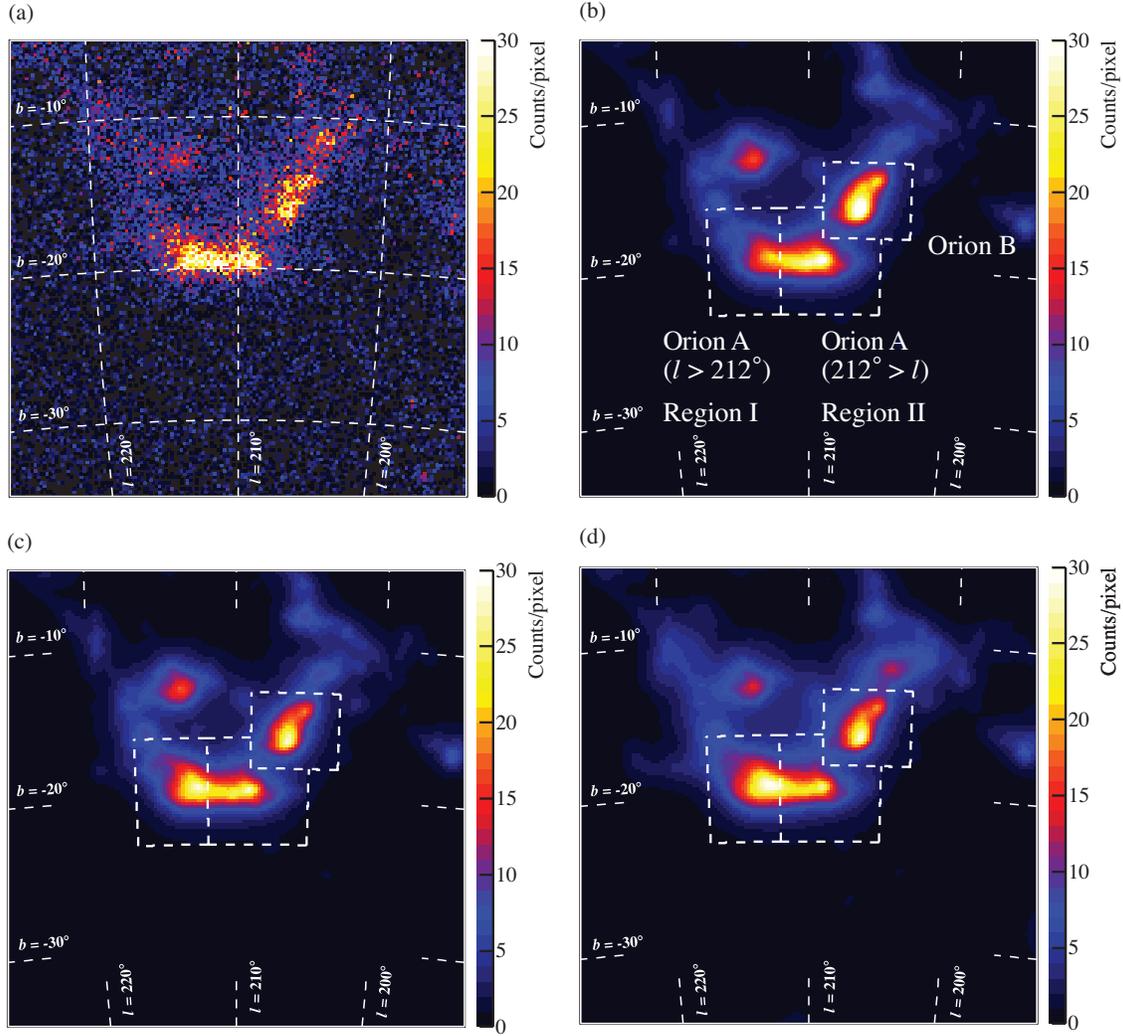}
\caption{ 
(a) Gamma-ray count distribution in the ROI
after subtracting the fitted ``background''
that is the sum of the \HI, IC, point-source, and isotropic components.
(b) The fitted model map obtained by assuming one common \Xco\ for the ROI
(\Hmol-template-1).
Dashed lines define the boundaries of the 3 Orion regions,
Orion A Region I, Region II and Orion B.
(c) Same as (b) but obtained by assuming 4 different \Xco\ for Orion A Region I,
Region II, Orion B, and elsewhere (\Hmol-template-2).
(d) Same as (b) but obtained by adding \EBVres\ to \Hmol-template-1 (\Hmol-template-3). }
\label{fig_model_map}
\end{figure}

We use \Hmol-template-1 as an approximation for the \Hmol\ gas distribution and fit Eq.~(\ref{eq_emission}) to determine the ``background''.
The energy-summed gamma-ray distribution after subtracting the ``background''
is shown in Fig.~\ref{fig_model_map}a and that of the \Wco-based model,
or the product of $\Sigma B_i$ in Eq.~(\ref{eq_emission}) and \Hmol-template-1,
is given in Fig.~\ref{fig_model_map}b.
The two count distributions are correlated pixel-by-pixel ($1\times 1$~deg$^2$)
in the 3 Orion regions in Fig.~\ref{fig_correlation}a.
We expect a good linear correlation between the two if \Wco\ is a good tracer 
of \Hmol. 

We note first that the correlation is fairly linear and gives a correlation coefficient\footnote{The correlation coefficient
is defined as
$\Sigma (x-\bar{x})(y-\bar{y})/\sqrt{\Sigma(x-\bar{x})^2\Sigma(y-\bar{y})^2}$}
of $0.93$.
We then note that the correlation significantly improves
if we separate the Orion clouds into
the 3 Orion regions, Orion A Region I (black solid line) and II (red dashed line),
and Orion B (blue dotted line).
The correlation coefficients for the 3 Orion regions are $0.98$, $0.96$, and $0.98$,
and the best-fit slopes are $0.72$, $0.99$, and $1.25$, respectively.

The large difference ($\sim40-60\%$) in the best-fit slope suggests that the mass column
density in Orion A and B cannot be simply derived using the same value of \Xco.
We find more gamma rays in Orion A Region I
per \Wco\ than in Orion A Region II and Orion B, suggesting \Xco\ is different
in the 3 Orion regions, or that some fraction of the \Hmol\ gas is not traced by \Wco\ provided a uniform CR density.
We explore these two possibilities by redefining the \Hmol-template.

\subsubsection{Spatial Fit with \Wco\ of 4 different \Xco\ values: \Hmol-template-2}
\label{subsubsec_HiCoDiv}
Based on the relation found between the spatial distributions of the
gamma-ray intensity associated with the \Hmol\ gas and 
the \Wco-based model (\Hmol-template-1),
we make a second template, \Hmol-template-2, that will delineate the \Hmol\ 
column density more faithfully.
In the template we divide the ROI into 4 regions, the 3 Orion regions and the rest of the ROI, and allow $B_i$ to be different in each region, or introduce 4 $B_i$'s.

The fitted results for $A_i$ (\HI)
and $B_i$ (\Hmol-template-2) in Eq.~(\ref{eq_emission}) are listed
in Table~\ref{ABCtable} after combining the highest 10 energy bins into 3 bins.
The gamma-ray count map is shown in Fig.~\ref{fig_model_map}c
is the sum of the 4 $B_i$'s multiplied with the corresponding components of
\Hmol-template-2. We note that the 3 Orion regions mix to some degree
through the \Fermi\ PSF. The correlation between the gamma-ray distribution 
associated with \Hmol\ and the \Hmol\ template
improved as shown in Fig.~\ref{fig_correlation}b:
the best-fit slopes for Orion A Region I, Region II and Orion B are $0.95$, $0.94$,
and $1.03$, respectively, while the correlation factors remain almost the same,
$0.98$, $0.99$, and $0.96$, respectively.

The \Xco\ for the 4 regions can be calculated directly as the ratio of
$B_i$ to $2A_i$ (the \Hmol/\HI\ method) or by extracting the gamma-ray emission
in the regions (the pionic method).
The results from the former are given in Table~\ref{tableXco} together with
those from the latter which will be described
in Subsection~\ref{subsec_spec_ana}.

\subsubsection{Spatial fit with \Wco\ and ``dark gas'': \Hmol-template-3}
\label{subsubsec_HiCoDark}

\citet{Grenier05} found that a significant fraction of local
diffuse gamma-ray emission observed by EGRET is not associated
with either \HI\ or \Wco, but rather with the dust map traced by thermal
infrared emission given by \citet{Schlegel98}.
The missing gas component is often referred to as the ``dark gas''.
Other LAT observations have found gamma rays associated with such ``dark gas''
\citep{Cassiopeia, ThirdQuadrant}. We note recent measurements of attenuation or
reddening of background stars have also detected gas concentrations
not traced well by \Wco\ \citep{Dobashi05, Rowles09, Dobashi11, Planck11}.

We make a third template, \Hmol-template-3, that can bring out the true gas distribution associated with the Orion clouds and enhance our understanding of the \Wco-to-\Hmol\ relation 
by introducing the ``dark gas''.
The new \Hmol\ template consists of \Hmol-template-1, or \Wco, and
a ``dark gas'' spatial template with a normalization factor for each.

Our ``dark gas'' template has been produced following the prescription
given by \citet{Grenier05} and referred to as \EBVres.
It is a residual map obtained by subtracting the best-fit
linear combination of \NHI\ and \Wco\ from the \EBV\ map
of \citet{Schlegel98} as described in \citet{ThirdQuadrant}.
Fig.~\ref{fig_template}c shows the \EBVres\ map around our ROI.
There is a problem with the color temperature correction of the map by \citet{Schlegel98}
around the OB associations in the Orion A and B clouds,
and thus \EBVres\ value is negative in these points. We masked out these pixels in the \EBVres\ map by setting the corresponding values to zero.

The results for $A_i$ (\HI) and $B_i$ (2 normalizations, one for \Wco\ and
one for the ``dark gas'') in Eq.~(\ref{eq_emission}) are listed after combining
the highest 10 energy bins
into 3 bins in Table~\ref{DarkGas_table}.
The distribution of the gamma-ray counts associated with \Hmol-template-3,
the sum of the counts associated with \Wco\ and the ``dark gas'',
is given in Fig.~\ref{fig_model_map}d.
The correlation between the extracted gamma-ray counts and the model counts improves
as shown in Fig.~\ref{fig_correlation}c, bringing the correlation coefficients
to $0.99$, $0.99$, $0.97$, and $0.98$, for Orion A Region I, Region II, Orion B
and the sum of the 3 regions, respectively.
The improvement in the correlation, or equivalently in the spatial fit, comes
from inclusion of \EBVres\, which has the largest contribution in
the Orion A Region I seen in Fig.~\ref{fig_template}c.

\subsubsection{Summary of the Spatial Fits}

The relative likelihoods of the spatial fits with Eq.~(\ref{eq_emission})
in the ROI are compared
among the 3 \Hmol-templates in Fig.~\ref{fig_ts} for the 22 energy bins.
The ``dark gas'' template (\Hmol-template-3) gives the best fit in almost all energy bins
and the 3-\Xco\ template (\Hmol-template-2) gives the second best result. The improvements relative to \Hmol-template-1 are statistically significant.

The residuals of the fits with the 3 templates in the ROI are given in Fig.~\ref{fig_model_residue}.
The rectangular boundaries of the 3 Orion regions shown
in Figs.~\ref{fig_model_map}b, c, d are replicated in the figure.
The residuals are significant within the Orion regions for
\Hmol-template-1 (Fig.~\ref{fig_model_residue}a) but not for the other 2 templates
(Fig.~\ref{fig_model_residue}b, c), which is consistent with the improvement
we saw in Fig.~\ref{fig_correlation}.
The difference in the residuals for \Hmol-template-2 and \Hmol-template-3 in the Orion regions is not significant
relative to the systematic uncertainty discussed in the next subsection.
We find that the large improvement \Hmol-template-3
has brought relative to \Hmol-template-2 in Fig.~\ref{fig_ts}
comes primarily from outside of the 3 Orion regions, especially in the Monoceros R2 region and in the northern region adjacent to the Orion~B:
the template adds ``dark gas'' in that part whereas the other templates only 
modify the 3 Orion regions.

The value of \Xco\ has been calculated by the \Hmol/\HI\ method by taking the ratio of
$B_i$ to $2A_i$ for the parts associated with \Wco\ in the \Hmol-templates
and listed in Table~\ref{tableXco}. In the pionic method of evaluating \Xco, however,
the pionic component must be extracted out of the gamma-ray spectrum
associated with the \Hmol-template as will be described
in Subsection~\ref{subsec_spec_ana}.
We will discuss the systematic errors in evaluating \Xco\
and possible interpretations of the results in Section~\ref{sec_dis}.

\subsection{Analyses of Spectra}
\label{subsec_spec_ana}
The spectra associated with the \HI\ and \Hmol-template-1,
with the \HI\ and \Hmol-template-2, and with the \HI\ and \Hmol-template-3
are obtained by assembling the fitted results for the respective templates,
$A_i$ and $B_i$, as shown in Figs.~\ref{fig_spec_fit_HiCoWoEMS},
\ref{fig_spec_fit_HiCoDivWoEMS}, and \ref{fig_spec_fit_HiCoDarkWoEMS},
respectively.  The spectra are fitted as a sum of the pionic and bremsstrahlung
components. The gamma-ray spectra associated with the spatial templates
(\HI, inverse Compton, isotropic, and sum of \Xco\ $\times$ \Wco)
are plotted for the 3 Orion regions in Fig.~\ref{fig_spec_all}a and b.
We analyze for the gamma rays associated
with the 3 \Hmol-templates in this subsection.

\subsubsection{Fit with Gamma-ray Emission Models}
\label{subsubsec_Spec_fit}
The spectral template of pionic gamma rays has been calculated by
convolving the gamma-ray inclusive cross-section for $p-p$ interaction
parameterized by \citet{Kamae06} and the CR proton spectrum predicted by GALPROP
at the Orion clouds.\footnote{In GALDEF 54\_77Xvarh7S, the CR proton flux was
artificially multiplied by $1.15$ to reproduce
gamma-ray observations by \Fermi.
The factor originates from the underestimate of gamma-ray emissivity for He and heavier atoms in the interstellar medium (ISM) in GALPROP. Instead of using the $1.15$ correction factor, we combined the calculation by \citet{Gaisser92} for contributions from CR He and heavier atoms, and the calculation by \citet{Mori09} for heavier atoms in the ISM.
Hence the total gamma-ray emissivity per H atom is $1.70$ times
larger than that for $p-p$ collisions only. The difference between the total gamma-ray emissivity in the two literatures is $\sim 5$\%, which is taken into account in the systematic uncertainty. \label{footnoteHI}}
The proton flux is predicted in the Orion clouds
($R=8.8$ kpc, $Z=-0.14$~kpc) to be $\sim8$\% smaller than
that at the solar system ($R=8.5$ kpc, $Z=0.0$~kpc)
where the GALPROP proton spectrum has been determined by the CR data
taken at the Earth. The value at the Orion clouds is consistent with that 
determined using the gamma rays from
the local \HI\  \citep{LocalHI09}.
The good fit to the data seen Figs.~\ref{fig_spec_fit_HiCoWoEMS}, \ref{fig_spec_fit_HiCoDivWoEMS}, and
\ref{fig_spec_fit_HiCoDarkWoEMS} supports GALPROP's prediction
of CR spectral shape in the Orion region and the overall modeling of Eq.~(\ref{eq_emission}).

Bremsstrahlung emission induced by CR electrons interacting with gas is
calculated in GALPROP using recent bremsstrahlung calculations \citep[][and references therein]{Strong98, Strong00}.
The electron injection spectrum in our GALPROP calculation had been adjusted to
reproduce, approximately, the power-law index of the electron spectrum measured 
by the \Fermi\ LAT \citep{Electron09}. In addition, the normalization of the spectrum is adjusted to reproduce the LAT observed gamma-ray flux at a low-energy band.
In the spectral fits described below, we kept the
electron-to-proton ratio, or equivalently the bremsstrahlung-to-pion ratio, fixed to the
value given in GALPROP. When we refer to the gamma-ray emissivity per atom
or molecule, we do not differentiate the underlying processes, but rather the
sum of the bremsstrahlung and pionic contributions.

The spectral fit of the \HI\ component is reasonable for all 3 \Hmol\ templates ($\chi^2=17.7$, $9.9$, and $17.1$ for $/\mathrm{dof}=14$ respectively)
as shown in Figs.~\ref{fig_spec_fit_HiCoWoEMS}a, \ref{fig_spec_fit_HiCoDivWoEMS}a
and \ref{fig_spec_fit_HiCoDarkWoEMS}a.
Our pionic flux associated with \HI\ is consistent with
that obtained in the \Fermi\ study on the local interstellar gas \citep{LocalHI09}
as overlaid in Fig.~\ref{fig_spec_fit_HiCoDivWoEMS}a. We note however that
there may be a small offset between the two as will be discussed later.
The spectra associated with molecular clouds are also fitted well by the 3
\Hmol-templates as shown below.

The mass-to-\Wco\ ratio, \Xco, can be obtained by comparing the assumed pionic gamma-ray emissivity per H atom with the observed gamma-ray emissivity per \Wco\ as shown in Figs.~\ref{fig_spec_fit_HiCoWoEMS}b, \ref{fig_spec_fit_HiCoDivWoEMS}b--d and \ref{fig_spec_fit_HiCoDarkWoEMS}b. The former is calculated in the unit of $\mathrm{MeV}^{-1}\mathrm{s}^{-1}\mathrm{sr}^{-1}$, and the latter is measured in the unit of $\mathrm{MeV}^{-1}\mathrm{s}^{-1}\mathrm{sr}^{-1}(2\times10^{20}\XcoUnit)^{-1}$. Thus, $\Xco/2$ of the clouds is derived by dividing the latter by the former.

The results of the spectral fit for the \HI\ component are not used to determine
\Xco\ in the pionic method.  The fits to the
spectral components shown in Figs.~\ref{fig_spec_fit_HiCoWoEMS}a, \ref{fig_spec_fit_HiCoDivWoEMS}a
and \ref{fig_spec_fit_HiCoDarkWoEMS}a are only to check overall consistency
of our analyses. Their normalizations are consistent within the uncertainty 
in the \HI\ column density discussed in Section~\ref{sec_dis}.   

\subsubsection{Spectra obtained with \Hmol-template-1}
The fitted spectra are plotted as sums of pionic and bremsstrahlung emissions in Fig.~\ref{fig_spec_fit_HiCoWoEMS}a, b for the \HI\ spatial template and
the \Hmol-template-1 (Orion A Region I, II, and Orion B combined),
giving $\chi^2/\mathrm{dof}$ of $17.7/14$ and $20.2/14$, respectively.

We give the \Xco\ value obtained from the fitted pionic spectra in Table~\ref{tableXco}.
Since the fit is substantially poorer than those for \Hmol-template-2 and 3
(see Fig.~\ref{fig_ts}), the value should be taken just as a reference value.
For this reason we do not quote systematic errors in the table.

\subsubsection{Spectra obtained with \Hmol-template-2}
The fitted spectra are plotted as sums of pionic and bremsstrahlung emissions in Fig.~\ref{fig_spec_fit_HiCoDivWoEMS}b, c, d for Orion A Region I, II and Orion B,
giving $\chi^2/\mathrm{dof}$ of $14.0/14$, $18.5/14$, and $10.6/14$, respectively.
The \Xco\ values obtained for the 4 regions from the fitted pionic spectra are
given in Table~\ref{tableXco}.

The coefficient \Xco\ is significantly higher for Orion A Region I than
for other regions, consistent with the slopes obtained in
Subsection~\ref{subsec_spatial_fit} in the pixel-by-pixel correlation study. 
This also can be seen in the \Xco\ obtained with the \Hmol/\HI\ method.

We note that the fraction of the \HI\ component in the gamma-ray spectrum integrated in the 3 Orion regions is comparable to that associated with \Wco\ 
(see Fig.~\ref{fig_spec_all}a). This is because the solid angle subtended by the Orion molecular clouds is a small fraction of our 3 Orion regions in solid angle and the overall mass of atomic gas is greater.

\subsubsection{Spectra obtained with \Hmol-template-3}
The fitted spectra integrated over the \Wco\ and ``dark gas'' components are shown in Fig.~\ref{fig_spec_fit_HiCoDarkWoEMS}b, c.  We give \Xco\
for the ROI from the fitted pionic spectrum in Table~\ref{tableXco}.

The \Xco\ obtained in fits with the \Wco\ can be compared with
those obtained in similar analyses including the ``dark gas'' template:
$2.0 \times10^{20}$ (in the local arm), $1.9 \times 10^{20}$
(the Perseus arm) and $0.87 \times 10^{20}$ (the Gould Belt) in the same unit as above \citep{ThirdQuadrant,Cassiopeia}.

The spectrum associated with the ``dark gas'' component is similar in shape
to that associated with \Wco\ but about half as intense (Fig.~\ref{fig_spec_all}b).
The 2 spectral energy densities (SEDs)  become comparable in Orion A Region I as seen
in Fig.~\ref{fig_spec_all}c. The ``dark gas'' dominates over \Wco\
in the pixels near the high-longitude end of Orion A and eventually
\Wco\ diminishes in the pixels beyond them towards higher longitude.

Our \Xco\ measurements given in Table~\ref{tableXco} can be compared with those
determined using the gamma-ray flux from the Orion-Monoceros complex
measured with EGRET: $(1.35\pm0.15)\times 10^{20}\ \XcoUnit$ \citep{Digel99}.
We note there were no Galactic CR propagation models
such as GALPROP nor CR measurements as precise as are
available now: \Xco\ was determined by the $\Hmol/\HI$ method and it compares well with the single \Xco\ value of $1.36 \pm 0.02$ obtained with the H2-template-1.

\subsection{Total masses of Orion A and B}
\label{subsec_mass_ana}

The distance from the Sun to the Orion nebula (M42)
inside the Orion A has recently been measured by parallax
to be $389^{+24}_{-21}$~pc \citep{Sandstrom07},
$414\pm7$ pc \citep{Menten07}, $437\pm 19$ pc \citep{Hirota07},
and $419 \pm 6$~pc \citep{Kim08}.
We adopted $400$~pc as the distance to the Orion A and B
clouds and used the total pionic gamma-ray fluxes obtained above
to get the total masses of Orion A and B outside\footnote{We note that the spatial extent
of Orion B defined here is significantly different from that used in \citet{Wilson05}
because we are unable to separate Orion B from the complex cloud structures
behind due to the broad PSF of the LAT.}.

{\underline{Mass estimation using \Hmol-template-2:}}
\begin{eqnarray}
  M_\mathrm{A} &= (74.5\pm1.3) \times 10^3 \; M_{400} \nonumber \\
  M_\mathrm{B} &= (33.5\pm0.7) \times 10^3 \; M_{400} \nonumber
\end{eqnarray}
where
\begin{equation}
  M_{400} = \left ( \frac{d}{400\ \mathrm{pc}}\right )^2 \times M_{\sun},
\end{equation}
and $d$ is the distance to the clouds.
We will discuss the systematic uncertainties in the next section.

{\underline{Mass estimation using \Hmol-template-3:}}
Addition of the ``dark gas'' changes the estimation of the Orion A and B masses
by about 10\%.
\begin{eqnarray}
  M_\mathrm{A, W_\mathrm{CO}} &= (55.1\pm0.8)\times 10^3\ M_{400} \nonumber \\
  M_\mathrm{A, Dark} &= (27.6\pm0.7)\times 10^3\ M_{400} \nonumber \\
  M_\mathrm{B} &= (36.0\pm0.5) \times 10^3\ M_{400}. \nonumber
\end{eqnarray}
The total mass of Orion A ($\equiv M_\mathrm{A, W_\mathrm{CO}} + M_\mathrm{A, Dark}$) is $(82.7 \pm 1.1)\times 10^3 M_{400}$. The Orion A mass has been estimated by \citet{Wilson05},
assuming $\Xco=1.8 \times 10^{20}\ \XcoUnit$
\citep{Dame01}, to be $M_\mathrm{A} = 91.7 \times 10^3\ M_{400}$.
The mass has been estimated separately for Orion A Regions 1, 2, 3, and NGC~2149 
in \citet{Wilson05}. Our Orion A (Region I and II) includes their Regions 1, 2, and 3 
but overlaps only partially with NGC~2149. Considering the breadth of the PSF 
and the limited statistics of the data, we could not determine how much
of NGC~2149 overlaps our Orion Region I.
If we assume about one half of NGC~2149 is
in our Orion Region I and the systematic error introduced by this ambiguity
is  half of the NGC~2149 mass estimated by \citet{Wilson05}, the Orion A mass
to be compared becomes $M_\mathrm{A} = (86.3\pm 5.4) \times 10^3\ M_{400}$.
The Orion B region is more complex and such a comparison is very difficult.

\section{Discussion}
\label{sec_dis}

Although the Orion clouds lie away from the Galactic Plane and subtend relatively
small solid angle, many Galactic and extragalactic sources contribute to
the ROI through the large PSF of the \Fermi-LAT.

We have analyzed the observed data to extract the intensity associated
with the molecular clouds, the 3 Orion regions in particular,
by using the 3 \Hmol-templates made from
\Wco\ on the 3 different assumptions for each of the 22 energy bins.
The ratio of the normalization factors for \HI\
and \Hmol, $A_i/2 B_i$, gives the conversion factor of \Wco\ to the mass column
density, \Xco\ (the \Hmol/\HI\ method). For this, the \HI\ mass column density
must be well understood from the radiative transfer of the \HI\ line and the CR spectrum must be constant in the ROI.

In the second method (the pionic method), \Xco\ is determined by comparing
the observed pionic gamma-ray intensities with those expected from
the CR spectrum at the Orion clouds and the pionic gamma-ray production
cross-section. For this, we have to know the absolute CR spectrum and flux,
the instrument response function (IRF), and the pionic gamma-ray
production cross-section, in particular the pionic gamma-ray contribution 
from metals in CR and ISM.

In the subsections to follow, we evaluate uncertainties and
possible systematic errors in the analyses, especially in
evaluating \Xco\ in the 3 Orion regions.
We then summarize the results obtained in this paper
and present possible interpretations thereon.

\subsection{Possible Systematic Errors in the Analyses}
\label{subsec_error_ana}

Systematic errors that affect the correlation measurements
between gamma-ray intensities and \Wco\ are discussed in two categories:
the first one applies commonly to the 3 Orion regions and the second
affects the relation differently in the 3 regions.

\subsubsection{CR intensity at the Orion clouds}
Uncertainty in the fluxes and spectra of CRs, in particular those of protons,
can affect in both categories. The Galactic CR protons that produce pions
in our energy range remain in our Galaxy longer ($\sim 5 \times 10^7$~yrs) than
electrons ($\sim 7 \times 10^6$~yrs)  \citep{LeeSH11} 
and their flux variation within the Galaxy is believed to be predicted 
well by GALPROP. We note that the CR source distribution, the Galaxy size, and 
the CR diffusion coefficient are the important inputs to GALPROP. 
Using the CR spectrum measured at the Earth,
we have calculated the CR spectrum in the Orion region
for the 2 choices of the CR source distributions and the 3 choices
of Galactic halo heights (2, 4, and 10~kpc) used in
a GALPROP-based study by \citet{LeeSH11}.
The CR spectrum does not change more than $\sim 2$\% from the value used here
as long as it is constrained to the measurements at the Earth
and to reproduce the Galactic diffuse gamma-ray intensities measured
by the \fermi\ LAT (see \citet{LeeSH11}).
We also note that the gamma-ray spectrum
from the local \HI\ (typical distance $< 1$~kpc) is consistent with the CR proton flux
being within $\sim 10$\% of that at the Earth \citep{LocalHI09}.

CRs could be accelerated in the clouds and/or prevented from penetrating
into their cores by embedded magnetic field.
We first note that there are no strong non-thermal X-ray source nor radio SNR found
in the clouds \citep[][and references therein]{Feigelson02}. Therefore
no appreciable CR acceleration is likely to be taking place in the Orion clouds.
The good linear correlation between \Wco\ and gamma-ray intensity seen
in all 3 Orion regions (Fig.~\ref{fig_correlation}a) confirms
that the CRs effective in producing pions (kinetic energy $> 1$~GeV) are
penetrating well inside the higher-density parts of the clouds.

Based on these observations we assume that the CR flux in
the Orion region is $8$\% lower than that at the Earth
with possible systematic error of $\pm 10$\% due 
mostly to disagreement among recent CR measurements at the Earth and solar demodulation uncertainties.

Uncertainty in the CR flux at the Orion clouds contributes directly
to the systematic error in the pionic method but indirectly
in the \Hmol/\HI\ method. In the former, the absolute CR intensity is assumed to be
known while the CR intensity is assumed to be the same in the local \HI\ region
and the molecular clouds in the latter.

\subsubsection{Uncertainty in the instrument response functions}
The uncertainty in the absolute calibration of the LAT effective area 
can also introduce error of the first kind.
The effective areas were derived based on Monte Carlo studies of the LAT,
 checked against beam tests at accelerators
\citep{Calib09, Atwood09}. Comparisons between flight data and Monte Carlo studies have been made to quantify the systematic uncertainty in the effective area \citep{Vela09}.
At present, we estimate this systematic error to be 10\% at 100~MeV,
5\% at 500~MeV and 20\% at 20~GeV.

The systematic error in the absolute energy scale has been estimated as $+5/-10$\% \citep{Electron09}. We have refitted \Xco\ after
artificially shifting the energy scale by $+5$\% and by $-10$\%:
the number of pionic gamma rays changes less than $+1/-8$\%
for all 3 Orion regions with all 3 \Hmol\ templates. We include
this possible error due to the uncertainty in the energy calibration when assessing the
overall systematic error.

The pionic method is affected directly by the uncertainty in the instrument response
function while the \Hmol/\HI\ method is insensitive because it affects the
denominator and numerator similarly.

\subsubsection{Uncertainty in the spin temperature of \HI}
In converting the observed 21~cm line emission intensity \citep{Kalberla05}
to the \HI\ column density, \Ts\ was assumed to be 125~K.
The range of \Ts\ measured in the local \HI\ gas varies broadly between
$90$~K and $400$~K \citep[e.g., ][and references therein]{Mohan04a,Mohan04b}
while we have assumed a likely range for our ROI
to be between $90$~K and $250$~K.

We refitted the \Fermi\ data in the ROI with these two extreme \Ts\ values
with \Hmol-template 2 and 3. We then calculated
\Xco\ by dividing $B_i$ by $2 A_i$ in Eq.~(\ref{eq_emission}),
or by extracting the pion component in
the spectra. The deviations of \Xco\ from those obtained with \Ts\ of 125~K
are taken into account in the systematic errors given in Table~\ref{tableXco}.
The large systematic errors for \Xco\ on $B_i/2 A_i$ (Column 3)
enter via $2 A_i$ which depends on the absolute calibration
of the \HI\ gas density or \Ts\ in the local \HI.
The pionic method uses the product of the CR intensity and
$pp \rightarrow \gamma$ cross-section in place of $2 A_i$ and
is less directly affected by the uncertainty in \HI\ gas density
or \Ts\ of the local \HI, although the uncertainties can have a
small indirect effect through the overall spatial fitting.
This effect is much smaller than the overall systematic error and negligible. 
We note that there is some discrepancy between
the gamma-ray spectra associated with \HI\ in the ROI and the local \HI\
\citep{LocalHI09} as seen in Fig.~\ref{fig_spec_fit_HiCoDivWoEMS}. 

\subsubsection{Effect of overlapping point source candidates}
We have not included the 5 sources overlapping with the Orion clouds (Sec.~\ref{subsec_pointsrc}) because
they are all classified as ``potentially confused with interstellar
diffuse emission or perhaps spurious'' \citep{1st_year}.
To investigate their potential contribution we repeated the analysis
including these sources with the fluxes
and spectra listed in the First \Fermi\ LAT Catalog.
The fit with the pionic method gives the following \Xco\ in unit of \XcoUnit:
$(2.29\pm0.05)\times10^{20}$ for Orion Region I;
$(1.16\pm0.05)\times10^{20}$ for Orion Region II;
and $(1.24\pm0.04)\times10^{20}$ for Orion B.
They are $2$\%, $19$\%, and $8$\% less than those obtained
without these point source candidates. In the present study, we assume they
are artifacts and add $+0/-2$, $+0/-19$, and $+0/-8$\% to the overall systematic 
error in the 3 regions.

\subsubsection{Overall error}
\label{subsubsec_total_error}
For the \Hmol/\HI\ method, the uncertainty in
the \HI\ mass density ($\sim 20$\%) due mostly to the uncertainty in \Ts\
dominates the systematic error . Other contributions include
the overlapping ``c'' sources ($+0/-2$, $+0/-19$, and $+0/-8$\%) and 
variation in the CR intensity within
$\sim 1$~kpc or between \HI\ and the molecular clouds ($\pm 5$\%),
making the total systematic errors for the 3 Orion regions to $+25/-28$, 
$+25/-44$, and $+25/-33$\% as given in Column 3 of Table~\ref{tableXco}.

For the pionic method,
the overall systematic error in determining \Xco\ comes from the uncertainty
in the IRF including that due to the energy calibration uncertainty ($\pm10$\%),
unknown contributions of the overlapping
sources ($+0/-2$, $+0/-19$, and $+0/-8$\%),
uncertainty in the CR intensity ($\pm 10$\%), uncertainty
in the $pp$ pion production cross-section ($\pm 5$\%), and 
uncertainty in the contribution from heavier nuclei ($\pm 5$\%).
We conservatively quote the linear sum of these combinations 
as the possible systematic error
for the 3 Orion regions, which are $+30/-32$, $+30/-49$, and $+30/-38$\%,
as given in Column 5 of Table~\ref{tableXco}.

The systematic errors that can affect \Xco\ differently in the 3 Orion regions
are variation in the CR intensity within $\sim 1$~kpc ($\pm 5$\%) 
and the overlapping sources.
The overall error of this kind is conservatively estimated to be the linear sum
of the two, $+5/-7$, $+5/-24$, and $+5/-13$\%.

\subsection{Gamma-ray intensity and \EJH}
\label{subsec_crosscheck}

\begin{figure}
\epsscale{1.0}
\plotone{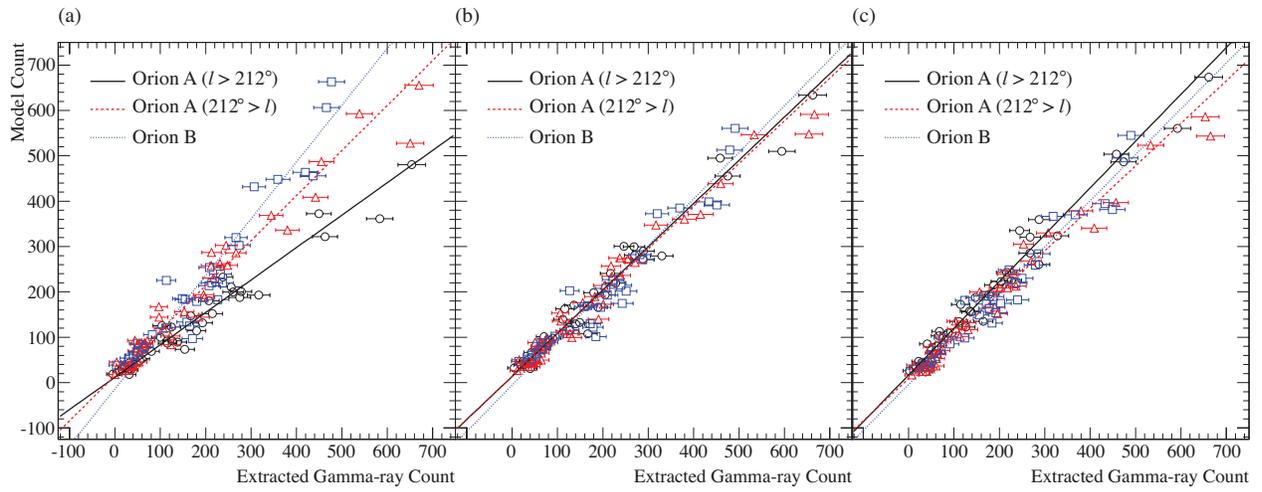}
\caption{ 
(a) Correlation between the gamma-ray count distribution shown
in Fig.~\ref{fig_model_map}a (the horizontal axis) and that fitted with
\Hmol-template-1 in Fig.~\ref{fig_model_map}b (the vertical axis) for all pixels
in the 3 Orion regions. Points represent pixels in Orion A Region I (black circles),
Region II (red triangle),
and Orion B (blue squares) with fitted lines black, red, and blue, respectively.
Error bars represent statistical errors in counts in pixels.
Same after replacing the vertical axis for that fitted with \Hmol-template-2 (b)
and for that fitted with \Hmol-template-3 (the sum of \Wco\ and \EBVres\ components) (c).}
\label{fig_correlation}
\end{figure}

\begin{figure}
\epsscale{0.7}
\plotone{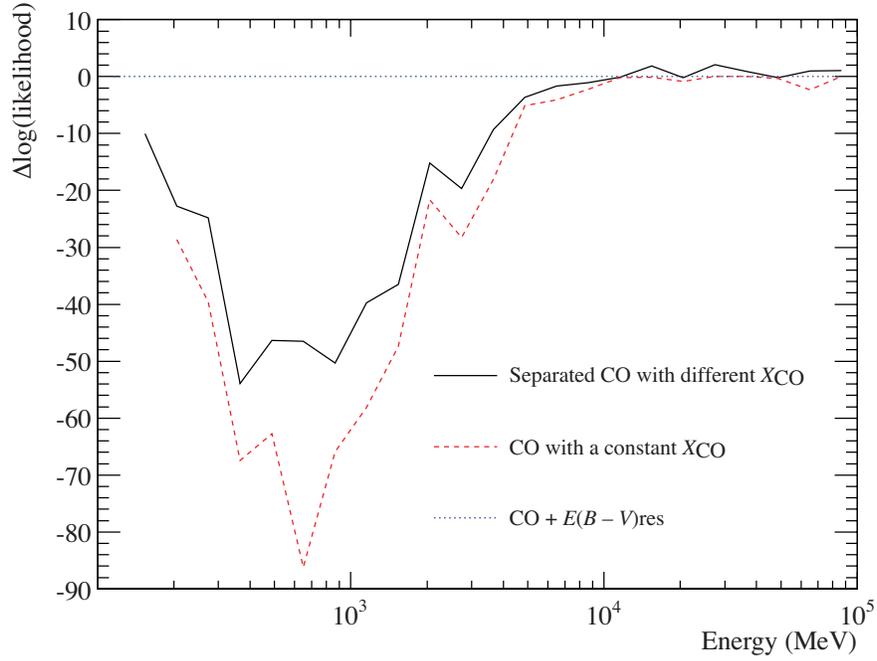}
\caption{
Difference in $\log(\mathrm{Likelihood})$ between the spatial fit using
\Hmol-template-3 (dotted line) and either that with \Hmol-template-1 (dashed line) or
that with \Hmol-template-2 (solid line) in the ROI for the 22 energy bins. Note that 
the lines are drawn between the data points only to guide the eye. }
\label{fig_ts}
\end{figure}

\begin{figure}
\epsscale{1.0}
\plotone{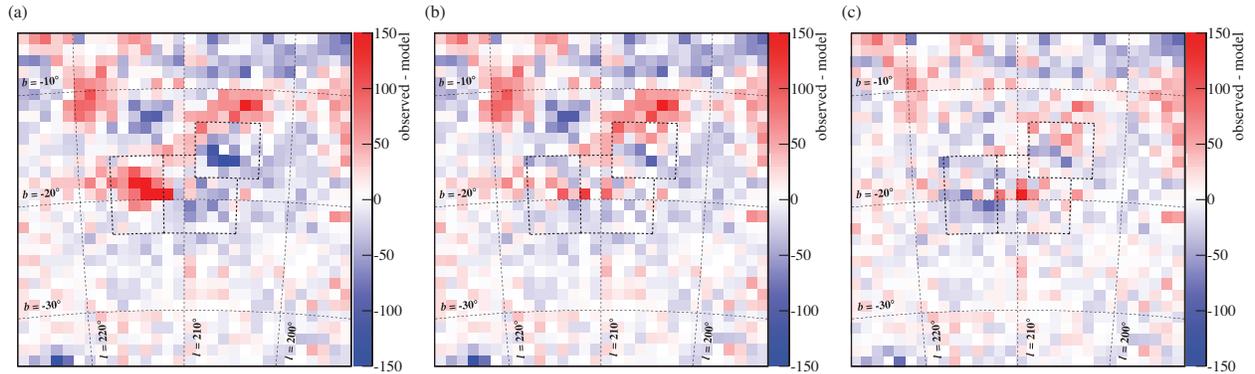}
\caption{Residue in the energy-summed gamma-ray counts of the spatial fit
with \Hmol-template-1 (a),
\Hmol-template-2 (b), and \Hmol-template-3 (c), binned in $1\times 1$~deg$^2$ pixels.
The black dotted lines show the boundaries of the 3 regions, Orion A Region I, II,
and Orion B.}
\label{fig_model_residue}
\end{figure}

\begin{figure}
\epsscale{1.0}
\plotone{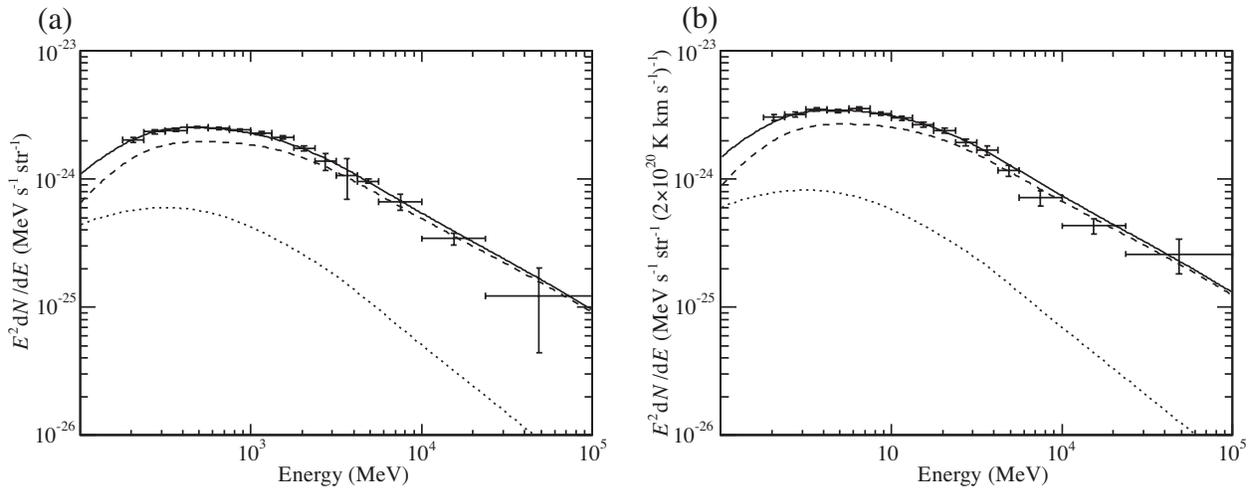}
\caption{Spectral energy densities (SED) associated with local
\HI\ ($\Ts=125\ \rm{K}$ assumed) (a) and
that associated with \Hmol-template-1 (b).
The lines are: total (solid),
bremsstrahlung (dotted) and pion decay (dashed). The CR spectral shape
and electron-to-proton ratio at the Orion clouds were fixed to those used by GALPROP. 
The vertical axes are normalized to the column density of \HI\ in unit of $1 \mathrm{cm}^{-1}$ for (a) and to
$2\times\Xco$ in unit of $10^{20}\rm{\ cm^{-2}(K\ km\ s^{-1})^{-1}}$ for (b).
The energy bins between No.13 and No.22 are combined to wider energy bins.
Vertical bars represent statistical errors.
Note that the spectral fit to \HI\ is not
used in evaluating \Xco.}
\label{fig_spec_fit_HiCoWoEMS}
\end{figure}

\begin{figure}
\epsscale{1.0}
\plotone{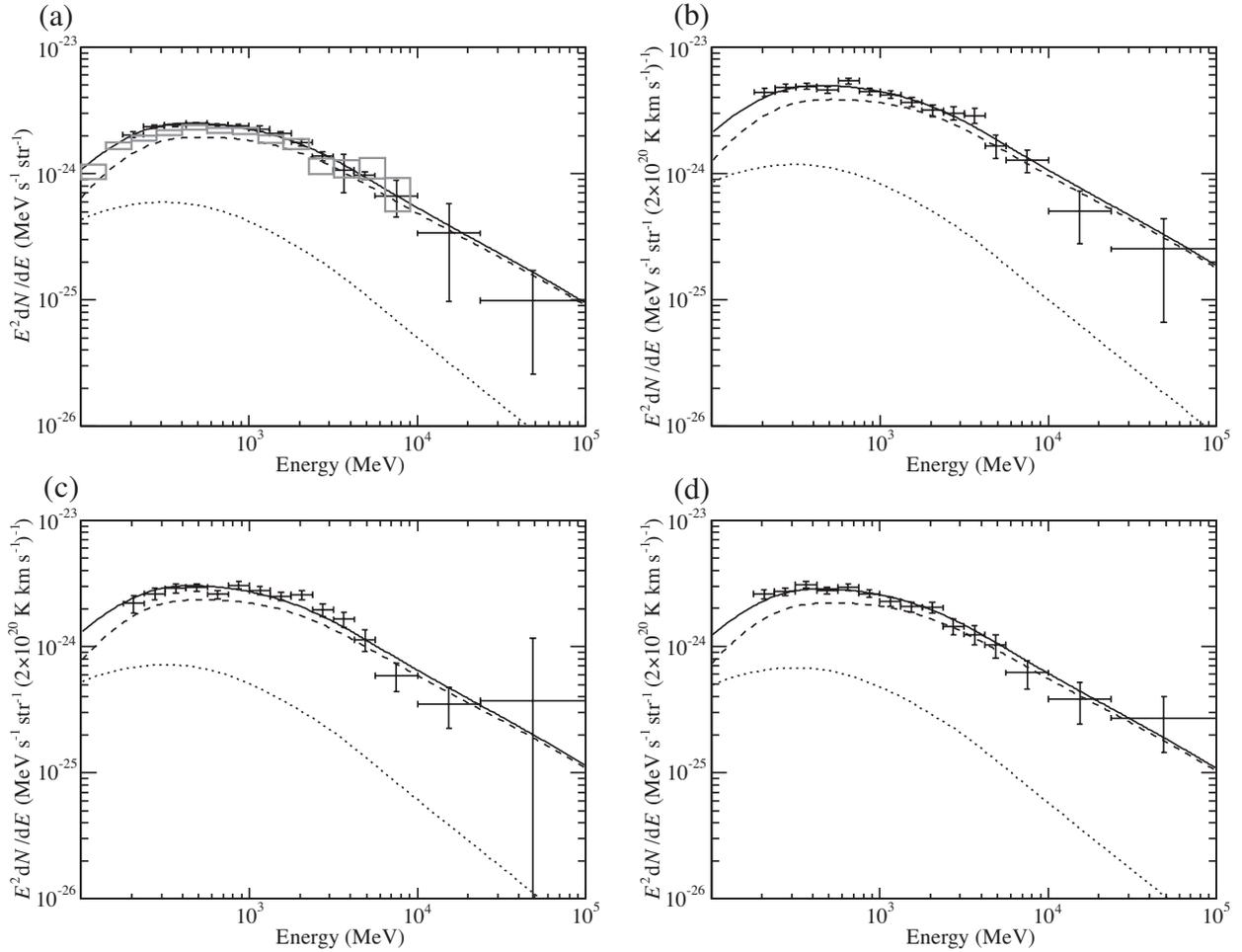}
\caption{Spectral energy density (SED) associated with local
\HI\ ($\Ts=125\ \rm{K}$ assumed) (a), Orion A Region I (b), Region II (c), and Orion B (d)
for the fit with \Hmol-template-2.
The corresponding SED obtained for the local \HI\ \citep{LocalHI09}
is shown by gray squares in (a). The assumption about the CR,
the line legends, and the vertical axis units are the same as in Fig.~\ref{fig_spec_fit_HiCoWoEMS}.}
\label{fig_spec_fit_HiCoDivWoEMS}
\end{figure}

\begin{figure}
\epsscale{1.0}
\plotone{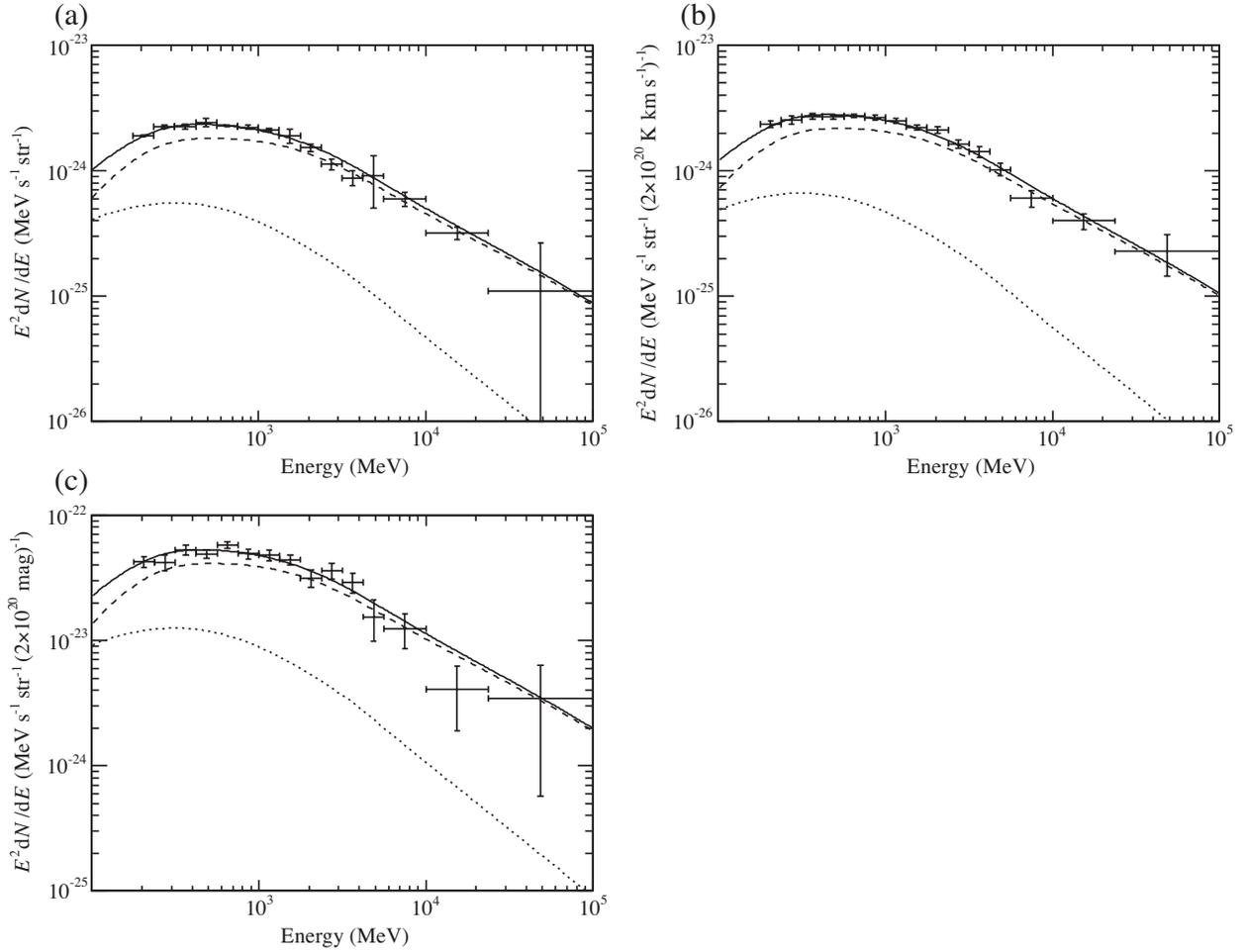}
\caption{SED associated with local
\HI\ ($\Ts=125\ \rm{K}$ assumed) (a),
that associated with \Wco\  (b), and that associated with \EBVres\ (c)
obtained with \Hmol-template-3. The line legends and vertical axis units are the same as in
Fig.~\ref{fig_spec_fit_HiCoWoEMS}.}
\label{fig_spec_fit_HiCoDarkWoEMS}
\end{figure}

\begin{figure}
\epsscale{1.}
\plotone{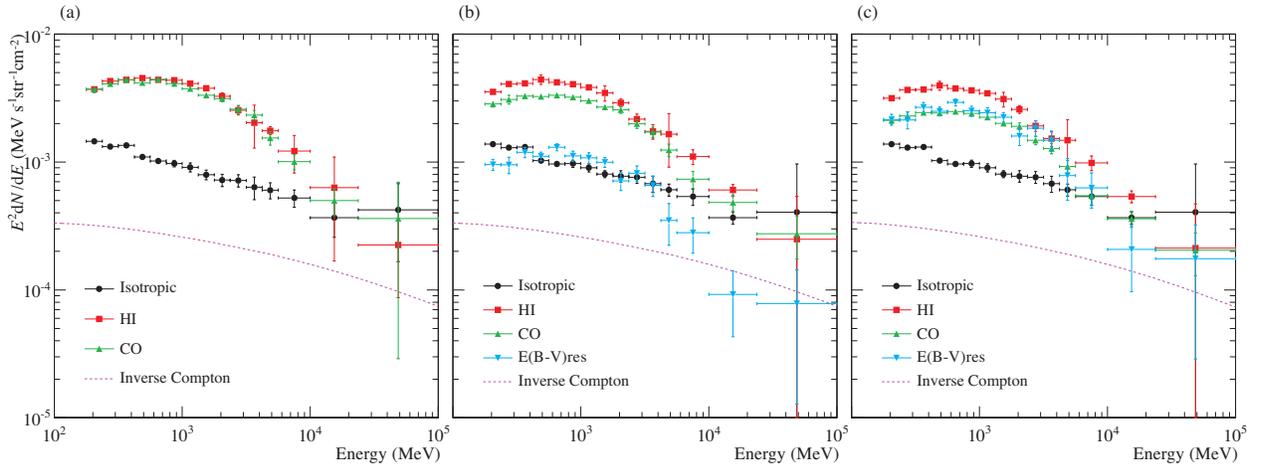}
\caption{
Gamma-ray spectra spatially associated with two \Hmol\ templates in the 3 Orion regions marked in Fig.~\ref{fig_model_map}b: (a) the sum of the 3 regions obtained
with \Hmol-template-2;
(b) the sum of the 3 regions with \Hmol-template-3; (c) Orion A Region I obtained
with \Hmol-template-3. Black circles show the isotropic component, red squares \HI, green upward triangles CO, and purple dashed line the inverse Compton. Blue downward triangles in (b) and (c) represent the spectra associated with \EBVres.}
\label{fig_spec_all}
\end{figure}

The line-of-sight visual attenuation, \Av, are often used
as a gas-mass tracer in theory-based studies of the CO
fraction in all molecules including carbon and hydrogen  
\citep[e.g.,][and references therein]{Burgh10, Wolfire10,Glover10}.
To calibrate crudely our mass column density with \Av\ used in these theory-based 
analyses, we have related the gamma-ray counts on the horizontal axes
of Fig.~\ref{fig_correlation} and \EJH\ in the
3 Orion regions measured by \citet{Dobashi11}. 
We note that the atomic and molecular components are assumed to be contained 
within a fixed length (e.g. ~20pc) along the line-of-sight in the theory-based analyses while 
the components are measured as column densities integrated over unknown 
lengths along the line-of-sight in observations. Moreover
\EJH\ is known to trace the \Hmol\ gas but also pick up some \HI\ gas 
through dust mixed with it. 
Hence the cross-calibration works at best crudely and only in the regions of clouds 
where the \Hmol\ longitudinal distribution is well confined and 
the \Hmol\ volume density dominates over that of \HI. Despite these uncertainties, 
it is important that our measurements be compared with theory-based analyses.

We found good linear relations for the pixels with high gamma-ray counts 
($>300$ per deg$^2$) in all 3 Orion regions and could correlate 
the gamma-ray count scale on the horizontal axes
of Fig.~\ref{fig_correlation} to \EJH\ assuming $\Av = R_{V-EJH} \times \EJH$.
The $R_{V-EJH}$ has been determined observationally and its value ranges
between 7.8 \citep{Dobashi11} to 10.9 \citep{Cardelli89}.
The highest point in our count map is $\sim 700$ per pixel in Orion A Region II 
where \Hmol\ concentration is highest and the corresponding value of \Av\ is 
$\sim 5$ when averaged over $1\times 1$~deg$^2$ pixels for an assumed value of  $R_{V-EJH}=7.8$. So $\Av = 5$ on the horizontal axes of Figs.~5 and 6 
in \citet{Glover10} corresponds crudely to $\sim 700$ counts per pixel 
assuming \Hmol\ is well confined (e.g., to $\sim 20$~pc) along the line-of-sight. 

\subsection{Summary of the Results}\label{subsec_summary}
The results obtained in the present work are significant beyond the estimated systematic
errors. They are:
\begin{enumerate}
\item Linearity holds between mass density associated with the Orion clouds and \Wco:
As discussed in Subsection~\ref{subsec_spatial_fit} and shown in
Fig.~\ref{fig_correlation}, our results
suggest that CRs penetrate to all translucent part of the clouds.
Possible shielding of CRs discussed in \citet{aharonian01}
does not apply to most parts of the Orion clouds.
\item The \Xco\ factors calculated with the pionic method and
with the \Hmol/\HI\ method differ by $\sim 15$\% but agree within
the estimated systematic error (Table~\ref{tableXco}). The difference 
can be explained by uncertainties in the column densities of \HI\ and calculation of gamma-ray emissivity per \HI\ atom.
\item The \Xco\ factor obtained with the \Hmol-template-2 is found to be larger by $\sim 40-60$\%
in Orion A Region I than Orion A Region II and Orion B for the two methods.
The difference is much larger than the systematic error that can affect the 
\Xco\ factor differently in the 3 Orion regions (Table~\ref{tableXco}).
\item In the ``dark gas'' scenario, the added ``dark gas'' accounts for the majority 
of the gas not traced by \Wco. One \Xco\ factor can then 
describe the \Wco-traced \Hmol\ distribution in the ROI. 
\end{enumerate}

\subsection{Interpretation of our results on \Xco}
\label{subsec_interpretation}
Historically the relation between \NHmol\ to \Wco\ has been considered to
depend on the environment around the molecular cloud. The
environmental factors discussed in the literature are:
\begin{description}
\item[Metallicity:] This possibility has been discussed in the literature since the late 1980's
\citep[e.g.,][]{Elmegreen89,Bolatto99}.  According to an empirical formula proposed to
relate $\Xco$ to [O/H] \citep{Wilson95, Arimoto96}, the metallicity must be
$\sim 2$ times higher in Orion A Region I to account for the observed difference in \Xco\
between Region I and II, which is unlikely according to Galactic-scale
measurements \citep[e.g.][]{Esteban05}.
We note that metallicity is generally considered to be an important
environmental factor influencing the \Hmol-to-\HI\ ratio.

\item[Overlapping \HI\ clumps:] Compact \HI\ clouds with angular diameters
of $1-2$deg have been found in various
Galactic locations \citep[e.g.][]{Braun86, Kavars03, LeeJJ08}.
A new reanalysis of the LAB \HI\ survey shows no such
concentration detected at the sensitivity level of the present study overlapping with
the Orion A and B \citep{Kalberla10}.

\item[Low density \Hmol\ not traced well by \Wco:] Existence of diffuse \Hmol\ gas
not traced well by \Wco\ has been discussed in the literature cited
in Section~\ref{sec_intro} and \Fermi\ analyses are bringing
the discussion to a quantitative level \citep{ThirdQuadrant}.
We refer to the following recent works on the \Hmol\ and CO fractions and
try to interpret our results:
  \begin{itemize}
    \item \citet{Burgh10} have studied the fractions based on {\it{Hubble Space
             Telescope}} observations and characterized the \Xco\ dependency on \NHmol.
    \item \citet{Wolfire10} have studied chemical composition of a model cloud
             theoretically and found that CO becomes depleted because of photodissociation in the periphery where
             the gas density decreases.
    \item \citet{Glover10} have studied the time-dependent \Hmol\ and CO fractions in clouds through
             computer simulations and found \Xco\ increases sharply where \NHmol\
             decreases for $\Av<3.5$.
  \end{itemize}

All of the above studies predict that the CO/(total C) fraction drops as
the \Hmol\ column density decreases, as toward the periphery of Orion A and B.
However the \Wco-to-\Hmol\ relation and the abundance of \Hmol-without-CO gas may be more complicated. For example,
\citet{Ikeda02} found that $N(\mbox{\ion{C}{1}})/N(\mathrm{CO})$
increases to high values along all of the peripheries whereas we find Region~I of Orion A to be more abundant in CO-depleted gas than Region~II.
The prediction that \Xco\ increases sharply in regions $\Av<3.5$ by
\citet{Glover10} is consistent with our finding that  the ``dark gas'' is concentrated
in the high-longitude end of Orion A where \Wco\ becomes low.
\end{description}

\section{Conclusion}
\label{sec_conclusion}
We have reported on the first 21 months' observations of
Orion A and B with the \textit{Fermi Gamma-ray Space Telescope}
in the energy band between $\sim178$~MeV and $\sim100$~GeV.
We have measured the mass column density distribution within the clouds
at the angular scale of the instrument PSF using the $pp \rightarrow \gamma$
production cross-section accurately calibrated at accelerators as well as 
using the gamma-ray emissivity of the local \HI\ gas.  We found with the pionic method
that a linear relation holds between mass density and \Wco\
with $\Xco =$ $2.34$, $1.43$, $1.35$ $\times10^{20}\ \XcoUnit$ with a
systematic uncertainty of $+5/-7$, $+5/-24$, and $+5/-13$\% 
(relative in the 3 regions), and $+30/-32$, $+30/-49$, and $+30/-38$\% 
(absolute) for Orion A Region I, Region II, and Orion B, respectively. 
These values are consistent with the \Xco\ values determined with 
the more traditional \Hmol/\HI\ method ($\Xco = $ $1.97$, $1.20$, $1.14$ $\times10^{20}\ \XcoUnit $) within our overall systematic error.
This implies that Galactic CRs are penetrating into most parts of the clouds.
The analyses also included the ``dark gas'' \citep{Grenier05} not traced by CO or \HI.
We found that the gamma-ray flux associated with the ``dark gas'' spatial template
exceeds that associated with the \Wco\ template in Orion A Region I. The situation is reversed in Region II and in Orion B.
This is generally consistent with the fit finding a higher \Xco\ value for Orion Region~I in the absence of the dark-gas template.

We have interpreted the increase in \Xco\ and ``dark gas'' fraction in Orion A Region I
in the light of recent studies of the relation between the \Hmol\
and CO fractions by \citet{Burgh10, Wolfire10, Glover10}.
\Xco\ is expected to increase rapidly as the gas column density
decreases to $\Av \sim 3.5$ or less \citep{Glover10}.
The mass column density we have measured in Region I corresponds
to $\Av<4$, close to the predicted threshold for onset of
the non-linearity predicted between \Wco\ and \NHmol. 
The mass column density drops further ($\Av < 2$) toward
the high Galactic longitude end of the Orion A where the gas becomes
``dark'' to \Wco, consistent with the predicted non-linear relation.

The \Fermi-LAT collaboration is continuing to reduce uncertainty in
the IRF, identify extended
gamma-ray sources, and improve the modeling of the Galactic-scale diffuse
gamma-ray emission. We expect the systematic uncertainties quoted
in subsection~\ref{subsec_error_ana} to be reduced significantly through
these efforts. The systematic uncertainty in the CR spectra and the \HI\
mass density also will be reduced when the data from new experiments
and surveys become available. The present analyses can then be updated
to a higher precision and the relation among \Wco\ and the gas mass
density characterized
further for various molecular clouds in the Galaxy.

\begin{center}
{\bf{\it{Acknowledgments}}}
\end{center}

The \textit{Fermi} LAT Collaboration acknowledges generous ongoing support
from a number of agencies and institutes that have supported both the
development and the operation of the LAT as well as scientific data analysis.
These include the National Aeronautics and Space Administration and the
Department of Energy in the United States, the Commissariat \`a l'Energie Atomique
and the Centre National de la Recherche Scientifique / Institut National de Physique
Nucl\'eaire et de Physique des Particules in France, the Agenzia Spaziale Italiana
and the Istituto Nazionale di Fisica Nucleare in Italy, the Ministry of Education,
Culture, Sports, Science and Technology (MEXT), High Energy Accelerator Research
Organization (KEK) and Japan Aerospace Exploration Agency (JAXA) in Japan, and
the K.~A.~Wallenberg Foundation, the Swedish Research Council and the
Swedish National Space Board in Sweden.

Additional support for science analysis during the operations phase is gratefully
acknowledged from the Istituto Nazionale di Astrofisica in Italy and the Centre National d'\'Etudes Spatiales in France.

\clearpage
\begin{deluxetable}{ccccc}
\tabletypesize{\scriptsize}
\rotate
\tablecolumns{5}
\tablecaption{Gamma-ray emissivity fitted with \Hmol-template-2}
\tablewidth{0pt}
\tablehead{
\colhead{Energy Range (MeV)} & \colhead{Emissivity per H \tablenotemark{a}} &
\multicolumn{3}{c}{Emissivity per \Wco\ \tablenotemark{b}} \\
\colhead{} & \colhead{} & \colhead{Orion Region I} & \colhead{Orion Region II}
& \colhead{Orion B}
}
\startdata
$178-237$ & $(4.81\pm0.26)\times10^{-29}$ & $(1.04\pm0.08)\times10^{-28}$ & $(5.20\pm0.84)\times10^{-29}$ & $(6.17\pm0.52)\times10^{-29}$\\
$237-316$ & $(3.15\pm0.10)\times10^{-29}$ & $(6.36\pm0.39)\times10^{-29}$ & $(3.50\pm0.37)\times10^{-29}$ & $(3.64\pm0.26)\times10^{-29}$\\
$316-422$ & $(1.81\pm0.04)\times10^{-29}$ & $(3.68\pm0.21)\times10^{-29}$ & $(2.16\pm0.18)\times10^{-29}$ & $(2.32\pm0.13)\times10^{-29}$\\
$422-562$ & $(1.05\pm0.02)\times10^{-29}$ & $(1.95\pm0.11)\times10^{-29}$ & $(1.24\pm0.09)\times10^{-29}$ & $(1.17\pm0.07)\times10^{-29}$\\
$562-750$ & $(5.72\pm0.12)\times10^{-30}$ & $(1.29\pm0.07)\times10^{-29}$ & $(6.15\pm0.46)\times10^{-30}$ & $(6.96\pm0.41)\times10^{-30}$\\
$750-1000$ & $(3.20\pm0.08)\times10^{-30}$ & $(5.97\pm0.37)\times10^{-30}$ & $(4.08\pm0.27)\times10^{-30}$ & $(3.50\pm0.24)\times10^{-30}$\\
$1000-1334$ & $(1.69\pm0.09)\times10^{-30}$ & $(3.16\pm0.23)\times10^{-30}$ & $(2.08\pm0.15)\times10^{-30}$ & $(1.70\pm0.14)\times10^{-30}$\\
$1334-1778$ & $(8.75\pm0.30)\times10^{-31}$ & $(1.55\pm0.13)\times10^{-30}$ & $(1.06\pm0.08)\times10^{-30}$ & $(8.71\pm0.80)\times10^{-31}$\\
$1778-2371$ & $(4.19\pm0.25)\times10^{-31}$ & $(7.49\pm0.77)\times10^{-31}$ & $(6.08\pm0.53)\times10^{-31}$ & $(4.82\pm0.49)\times10^{-31}$\\
$2371-3162$ & $(1.83\pm0.14)\times10^{-31}$ & $(4.01\pm0.47)\times10^{-31}$ & $(2.60\pm0.29)\times10^{-31}$ & $(1.92\pm0.27)\times10^{-31}$\\
$3162-4217$ & $(7.97\pm2.72)\times10^{-32}$ & $(2.16\pm0.29)\times10^{-31}$ & $(1.23\pm0.17)\times10^{-31}$ & $(9.34\pm1.56)\times10^{-32}$\\
$4217-5623$ & $(4.07\pm0.29)\times10^{-32}$ & $(6.98\pm1.46)\times10^{-32}$ & $(4.77\pm0.95)\times10^{-32}$ & $(4.33\pm0.90)\times10^{-32}$\\
$5623-10000$ & $(1.19\pm0.38)\times10^{-32}$ & $(2.27\pm0.47)\times10^{-32}$ & $(1.05\pm0.27)\times10^{-32}$ & $(1.10\pm0.28)\times10^{-32}$\\
$10000-23714$ & $(1.42\pm1.01)\times10^{-33}$ & $(2.12\pm0.95)\times10^{-33}$ & $(1.48\pm0.53)\times10^{-33}$ & $(1.61\pm0.59)\times10^{-33}$\\
$23714-100000$ & $(4.16\pm3.06)\times10^{-35}$ & $(1.07\pm0.79)\times10^{-34}$ & $(1.57\pm3.34)\times10^{-34}$ & $(1.14\pm0.53)\times10^{-34}$
\enddata
\label{ABCtable}
\tablecomments{Errors are statistical only}
\tablenotetext{a}{$\mathrm{MeV}^{-1}\mathrm{s}^{-1}\mathrm{sr}^{-1}$ per H atom}
\tablenotetext{b}{$\mathrm{MeV}^{-1}\mathrm{s}^{-1}\mathrm{sr}^{-1}(2\times10^{20}\XcoUnit)^{-1}$}
\end{deluxetable}

\begin{deluxetable}{lllll}
\tablecolumns{5}
\tablewidth{15cm}
\tablecaption{\Xco\ obtained on \Hmol-template-1, 2, and 3}
\tablehead{\colhead{Region}	& \colhead{\Xco$^a$ on $B/2A$}	& \colhead{Sys. error$^b$ (\%)} &
\colhead{\Xco$^a$ on pion} & \colhead{Sys. error$^c$ (\%)} }
\startdata
\multicolumn{5}{l}{\Hmol-template-1} \\
\hline
Entire ROI & $1.36\pm 0.02_\mathrm{stat}$ & NA & $1.63\pm 0.02_\mathrm{stat}$ & NA \\
\hline
\multicolumn{5}{l}{\Hmol-template-2} \\
\hline
Orion~A~Region~I & $1.97\pm0.05_\mathrm{stat}$ & $+25/-28$ &
$2.34\pm0.05_\mathrm{stat}$ & $+30/-32$ \\
Orion~A~Region~II & $1.20\pm0.03_\mathrm{stat}$ & $+25/-44$ &
$1.43\pm0.04_\mathrm{stat}$ & $+30/-49$ \\
Orion~B & $1.14\pm0.03_\mathrm{stat}$ & $+25/-33$ &
$1.35\pm0.03_\mathrm{stat}$ & $+30/-38$ \\
Elsewhere & $1.43\pm0.04_\mathrm{stat}$ & NA$^c$ &
$1.69\pm0.04_\mathrm{stat}$ & NA$^d$ \\
\hline
\multicolumn{5}{l}{\Hmol-template-3} \\
\hline
Entire ROI &  $1.21\pm0.02_\mathrm{stat}$ & $+25/-37$\%$^e$ &
$1.32\pm0.02_\mathrm{stat}$ & $+30/-40$$^e$ \\
\enddata
\label{tableXco}
\tablenotetext{a}{In unit of $10^{20}$ \XcoUnit.}
\tablenotetext{b}{The systematic error is discussed in
Subsection~\ref{subsec_error_ana}: it comes from a
combination of uncertainties in the \HI\ spin temperature and in the
fitting process. The systematic errors which may apply differently
to the 3 Orion regions are $+5/-8$, $+5/-24$, and $+5/-13$\%, respectively.}
\tablenotetext{c}{The systematic error is discussed in
Subsection~\ref{subsec_error_ana}. The systematic errors are the same as b.}
\tablenotetext{d}{We have not attempted to estimate systematic error outside of
the Orion regions in this study.}
\tablenotetext{e}{The average of the systematic errors estimated for
the 3 Orion regions.}
\end{deluxetable}

\begin{deluxetable}{cccc}
\tabletypesize{\scriptsize}
\tablecolumns{4}
\tablecaption{Gamma-ray emissivity fitted with \Hmol-template-3}
\tablewidth{0pt}
\tablehead{
\colhead{Energy Range (MeV)} & \colhead{Emissivity per H atom\tablenotemark{a}} & \colhead{Emissivity per \Wco\tablenotemark{b}}& \colhead{Emissivity per \EBVres\tablenotemark{c}}
}
\startdata
$178-237$ & $(4.51\pm0.08)\times10^{-29}$ & $(5.56\pm0.31)\times10^{-29}$ & $(1.00\pm0.10)\times10^{-27}$\\
$237-316$ & $(2.99\pm0.08)\times10^{-29}$ & $(3.39\pm0.27)\times10^{-29}$ & $(5.60\pm0.83)\times10^{-28}$\\
$316-422$ & $(1.68\pm0.07)\times10^{-29}$ & $(2.03\pm0.11)\times10^{-29}$ & $(3.95\pm0.36)\times10^{-28}$\\
$422-562$ & $(1.02\pm0.08)\times10^{-29}$ & $(1.13\pm0.05)\times10^{-29}$ & $(2.07\pm0.16)\times10^{-28}$\\
$562-750$ & $(5.39\pm0.08)\times10^{-30}$ & $(6.51\pm0.21)\times10^{-30}$ & $(1.37\pm0.09)\times10^{-28}$\\
$750-1000$ & $(2.97\pm0.09)\times10^{-30}$ & $(3.54\pm0.16)\times10^{-30}$ & $(6.57\pm0.60)\times10^{-29}$\\
$1000-1334$ & $(1.58\pm0.05)\times10^{-30}$ & $(1.86\pm0.09)\times10^{-30}$ & $(3.57\pm0.35)\times10^{-29}$\\
$1334-1778$ & $(8.00\pm1.02)\times10^{-31}$ & $(9.37\pm0.43)\times10^{-31}$ & $(1.86\pm0.16)\times10^{-29}$\\
$1778-2371$ & $(3.64\pm0.25)\times10^{-31}$ & $(5.00\pm0.31)\times10^{-31}$ & $(7.45\pm1.19)\times10^{-30}$\\
$2371-3162$ & $(1.51\pm0.14)\times10^{-31}$ & $(2.19\pm0.17)\times10^{-31}$ & $(4.82\pm0.68)\times10^{-30}$\\
$3162-4217$ & $(6.56\pm0.89)\times10^{-32}$ & $(1.06\pm0.10)\times10^{-31}$ & $(2.18\pm0.40)\times10^{-30}$\\
$4217-5623$ & $(3.82\pm1.71)\times10^{-32}$ & $(4.31\pm0.49)\times10^{-32}$ & $(6.50\pm2.34)\times10^{-31}$\\
$5623-10000$ & $(1.06\pm0.14)\times10^{-32}$ & $(1.07\pm0.16)\times10^{-32}$ & $(2.20\pm0.68)\times10^{-31}$\\
$10000-23714$ & $(1.35\pm0.15)\times10^{-33}$ & $(1.68\pm0.24)\times10^{-33}$ & $(1.72\pm0.92)\times10^{-32}$\\
$23714-100000$ & $(4.62\pm6.52)\times10^{-35}$ & $(9.55\pm3.50)\times10^{-35}$ & $(1.46\pm1.22)\times10^{-33}$
\enddata
\label{DarkGas_table}
\tablecomments{Errors are statistical only}
\tablenotetext{a}{$\mathrm{MeV}^{-1}\mathrm{s}^{-1}\mathrm{sr}^{-1}$ per H atom}
\tablenotetext{b}{$\mathrm{MeV}^{-1}\mathrm{s}^{-1}\mathrm{sr}^{-1}(2\times10^{20}\XcoUnit)^{-1}$}
\tablenotetext{c}{$\mathrm{MeV}^{-1}\mathrm{s}^{-1}\mathrm{sr}^{-1}(2\times10^{20}\mathrm{mag})^{-1}$}
\end{deluxetable}

\clearpage

\end{document}